\documentclass[multphys,vecphys]{svmult}

\usepackage{makeidx}         
\usepackage{graphicx}        
\usepackage{multicol}        
\usepackage[bottom]{footmisc}

\makeindex             
\def\ga{\mathrel{\raise.3ex\hbox{$>$\kern-.75em\lower1ex\hbox{$\sim$}}}}
\def\la{\mathrel{\raise.3ex\hbox{$<$\kern-.75em\lower1ex\hbox{$\sim$}}}}
\def\beq{\begin{equation}}  
\def\eeq{\end{equation}}
\def\ohsq{\Omega_{\chi} h^2}
\def\m12{m_{1\!/2}}
\def\mst{m_{\tilde\tau_1}}

\def\st{{\widetilde \tau}_{\scriptscriptstyle\rm 1}}

\def\gev{{\rm \, Ge\kern-0.125em V}}
\def\ba{\begin{eqnarray}}
\def\ea{\end{eqnarray}}

\begin{document}

\title*{Dark Matter Candidates in Supersymmetric Models$^1$}
\author{Keith A. Olive}
\institute{William I.\ Fine Theoretical Physics Institute,\\
University of Minnesota, Minneapolis, MN~55455, USA
\texttt{olive@umn.edu}\\
{\footnotesize $^1$To be published in 
``Dark 2004", proceedings of 5th International Heidelberg Conference 
on Dark Matter in Astro and Particle Physics, eds. H.V. Klapdor-Kleingrothaus and R. Arnowitt.}}
%
%
\maketitle

\vspace*{-2.6in}
\rightline{hep-ph/0412054}
\rightline{UMN--TH--2332/04}
\rightline{FTPI--MINN--04/45}
\rightline{December 2004}
\vskip 2.0in

The status of the constrained minimal supersymmetric standard model (CMSSM) will 
be discussed in light of our current understanding of the relic density
after WMAP. A global likelihood analysis of the model is performed including 
data from direct Higgs searches, global fits to electroweak data, $b \to s \gamma$, 
the anomalous magnetic moment of the muon, as well as the cosmological relic 
density.  Also considered are models which relax and further constrain the CMSSM.
Prospects for dark matter detection in colliders and cryogenic detectors will be 
briefly discussed.

\section{Introduction}
\label{sec:1}
Supersymmetric models with conserved $R$-parity
contain one new stable particle which is a candidate for cold dark matter
(CDM) \cite{EHNOS}. 
There are very strong constraints, however, forbidding the existence of stable or
long lived particles which are not color and electrically neutral.
 The sneutrino \cite{snu} is one
possible candidate, but in the MSSM, it has been excluded as a dark matter
candidate by direct \cite{dir} and indirect \cite{indir} searches.  Another possibility is the
gravitino and is probably the most difficult to exclude. 
This possibility has been discussed recently in the CMSSM context \cite{gdm}.
I will concentrate on the remaining possibility in the MSSM, namely the
neutralinos.

 There are four neutralinos, each of which is a  
linear combination of the $R=-1$, neutral fermions \cite{EHNOS}: the wino
$\tilde W^3$, the partner of the
 3rd component of the $SU(2)_L$ gauge boson;
 the bino, $\tilde B$, the partner of the $U(1)_Y$ gauge boson;
 and the two neutral Higgsinos,  $\tilde H_1$ and $\tilde H_2$.
 In general, the
neutralino mass eigenstates can  be expressed as a linear combination
\begin{equation}
	\chi = \alpha \tilde B + \beta \tilde W^3 + \gamma \tilde H_1 +
\delta
\tilde H_2
\end{equation}
The solution for the coefficients $\alpha, \beta, \gamma$ and $\delta$
for neutralinos that make up the LSP 
can be found by diagonalizing the mass matrix
which depends on $M_1 (M_2)$ which are the soft supersymmetry breaking
 U(1) (SU(2))  gaugino mass terms, $\mu$, the supersymmetric Higgs
mixing mass parameter and the two Higgs vacuum expectation values, $v_1$ and $v_2$. 
One
combination of these is related to the $Z$ mass, and therefore is not a
free parameter, while the other combination, the ratio of the two vevs,
$\tan \beta$, is free.  

The most general version of the MSSM, despite its minimality in particles and
interactions contains well over a hundred new parameters. The study of such a model
would be untenable were it not for some (well motivated) assumptions.
These have to do with the parameters associated with supersymmetry breaking.
It is often assumed that, at some unification scale, all of the gaugino masses
receive a common mass, $m_{1/2}$. The gaugino masses at the weak scale are
determined by running a set of renormalization group equations.
Similarly, one often assumes that all scalars receive a common mass, $m_0$,
at the GUT scale. These too are run down to the weak scale. The remaining
supersymmetry breaking parameters are the trilinear mass terms, $A_0$,
which I will also assume are unified at the GUT scale,  and the bilinear
mass term
$B$. There are, in addition, two physical CP violating phases which will
not be considered here.

The natural boundary conditions at the GUT scale for the MSSM would
include
$\mu$ and $B$ in addition to
$m_{1/2}$,
$m_0$, and $A_0$. In this case, upon running the RGEs down to a low energy
scale and minimizing the Higgs potential, one would predict the values of $M_Z$, 
$\tan \beta$ (in addition to all of the sparticle masses).
Since $M_Z$ is known, it is more useful to analyze supersymmetric models
where $M_Z$ is input rather than output.  It is also common to treat
$\tan \beta$ as an input parameter. This can be done at the expense of 
shifting $\mu$ (up to a sign) and $B$ from inputs to outputs. 
This model is often referred
to as the constrained MSSM or CMSSM. Once these parameters are set, the
entire spectrum of sparticle masses at the weak scale can be calculated. 
In the CMSSM, the solutions for $\mu$ generally lead to a neutralino
which which very nearly a pure $\tilde B$.

\section{The CMSSM after WMAP}
\label{sec:2}

\begin{figure}[h]
\includegraphics[height=2.3in]{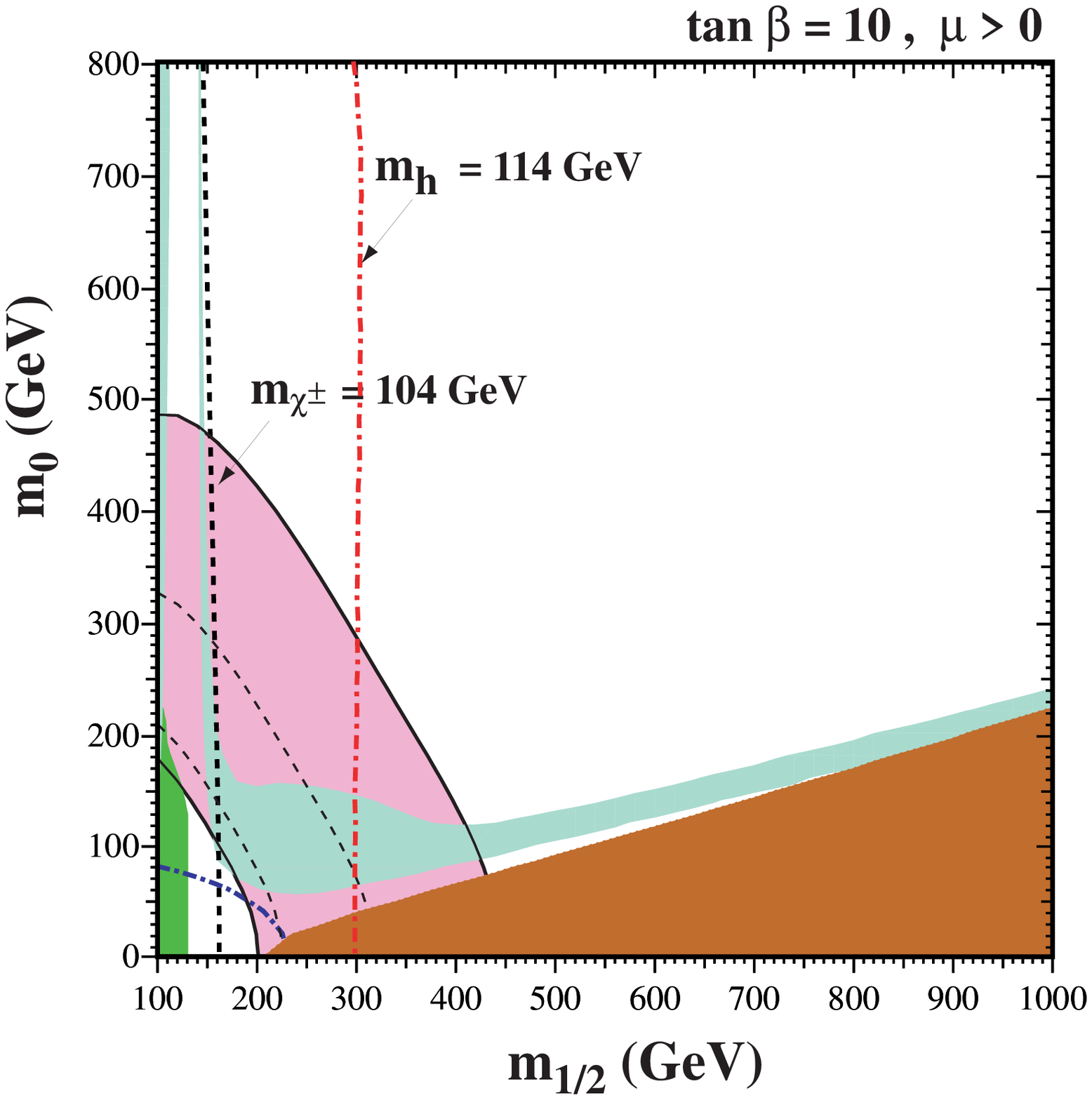}
\includegraphics[height=2.3in]{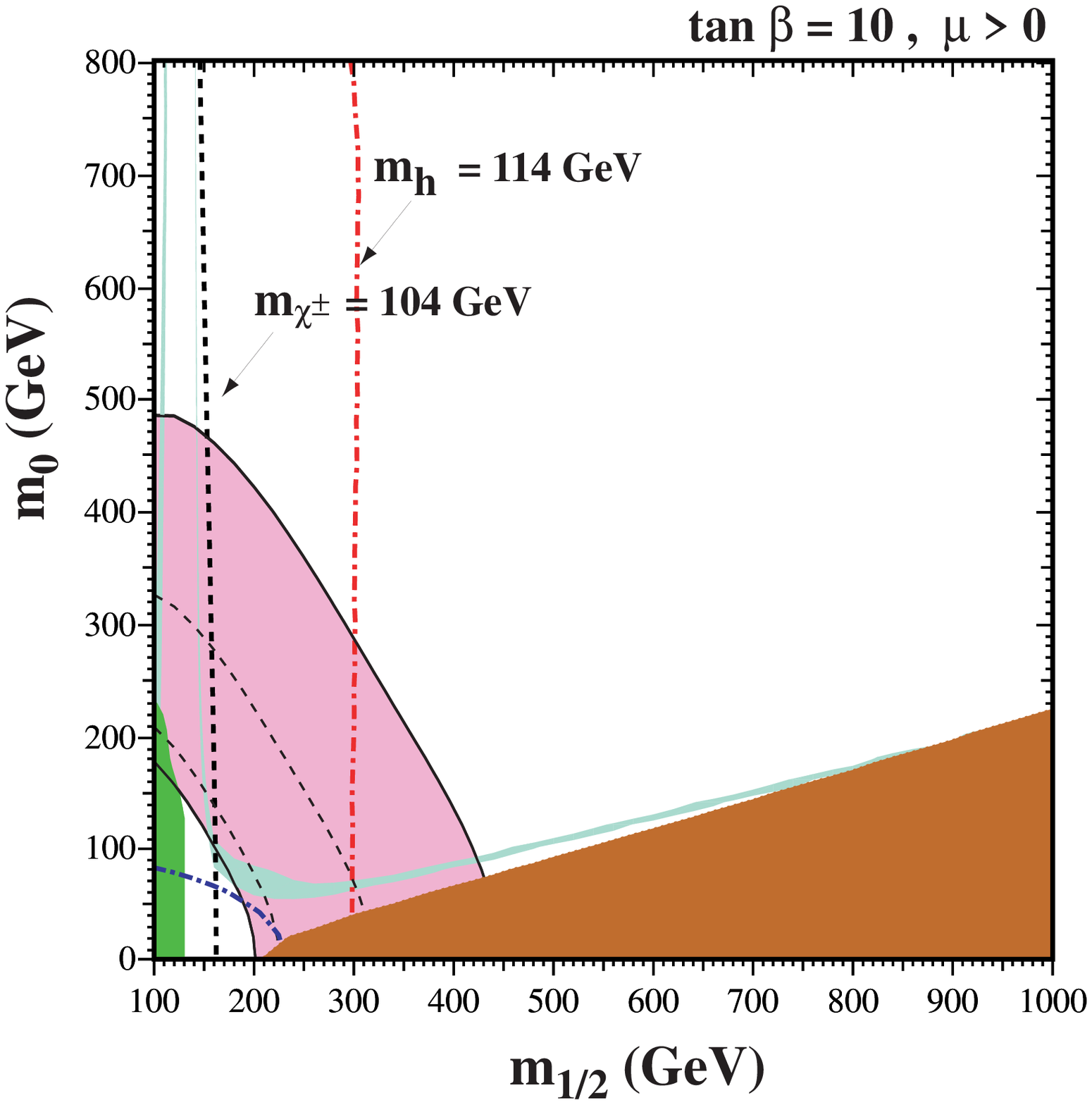}
\caption{\label{fig:UHM}
{\it The $(m_{1/2}, m_0)$ planes for  (a) $\tan \beta = 10$ and  $\mu > 0$, 
assuming $A_0 = 0, m_t = 175$~GeV and
$m_b(m_b)^{\overline {MS}}_{SM} = 4.25$~GeV. The near-vertical (red)
dot-dashed lines are the contours $m_h = 114$~GeV, and the near-vertical (black) dashed
line is the contour $m_{\chi^\pm} = 104$~GeV. Also
shown by the dot-dashed curve in the lower left is the corner
excluded by the LEP bound of $m_{\tilde e} > 99$ GeV. The medium (dark
green) shaded region is excluded by $b \to s
\gamma$, and the light (turquoise) shaded area is the cosmologically
preferred regions with \protect\mbox{$0.1\leq\ohsq\leq 0.3$}. In the dark
(brick red) shaded region, the LSP is the charged ${\tilde \tau}_1$. The
region allowed by the E821 measurement of $a_\mu$ at the 2-$\sigma$
level, is shaded (pink) and bounded by solid black lines, with dashed
lines indicating the 1-$\sigma$ ranges. In (b), the relic density is restricted to the
range $0.094 < \ohsq < 0.129$. }}
\end{figure}

For a given value of $\tan \beta$, $A_0$,  and $sgn(\mu)$, the resulting regions of 
acceptable relic density and which satisfy the phenomenological constraints
can be displayed on the  $m_{1/2} - m_0$ plane.
In Fig. \ref{fig:UHM}a,  the light
shaded region corresponds to that portion of the CMSSM plane
with $\tan \beta = 10$, $A_0 = 0$, and $\mu > 0$ such that the computed
relic density yields \mbox{$0.1<\ohsq<0.3$}.
At relatively low values of 
$m_{1/2}$ and $m_0$, there is a large  `bulk' region  which tapers off
as $\m12$ is increased.  At higher values of $m_0$,  annihilation cross sections
are too small to maintain an acceptable relic density and $\ohsq > 0.3$.
Although sfermion masses are also enhanced at large $\m12$ (due to RGE running),
co-annihilation processes between the LSP and the next lightest sparticle 
(in this case the $\st$) enhance the annihilation cross section and reduce the
relic density.  This occurs when the LSP and NLSP are nearly degenerate in mass.
The dark shaded region has $m_{\st}< m_\chi$
and is excluded.   Neglecting coannihilations, one would find an upper
bound of $\sim450\gev$ on $\m12$, corresponding to an upper bound of
roughly $200\gev$ on $m_{\tilde B}$.  
The effect of coannihilations is
to create an allowed band about 25-50 $\gev$ wide in $m_0$ for $\m12 \la
1400\gev$, which tracks above the $\mst=m_\chi$ contour \cite{efo}.

Also shown in Fig. \ref{fig:UHM}a are
the relevant phenomenological constraints.  
These include the limit on the chargino mass: $m_{\chi^\pm} > 104$~GeV \cite{LEPsusy}, 
on the selectron mass: $m_{\tilde e} > 99$~GeV 
 \cite{LEPSUSYWG_0101} and on the Higgs mass: $m_h >
114$~GeV \cite{LEPHiggs}. The former two constrain $m_{1/2}$ and $m_0$ directly
via the sparticle masses, and the latter indirectly via the sensitivity of
radiative corrections to the Higgs mass to the sparticle masses,
principally $m_{\tilde t, \tilde b}$. 
{\tt FeynHiggs}~\cite{FeynHiggs} is used for the calculation of $m_h$. 
The Higgs limit  imposes important constraints
principally on $m_{1/2}$ particularly at low $\tan \beta$.
Another constraint is the requirement that
the branching ratio for $b \rightarrow
s \gamma$ is consistent with the experimental measurements \cite{bsgex}. 
These measurements agree with the Standard Model, and
therefore provide bounds on MSSM particles \cite{gam,bsgth},  such as the chargino and
charged Higgs masses, in particular. Typically, the $b\rightarrow s\gamma$
constraint is more important for $\mu < 0$, but it is also relevant for
$\mu > 0$,  particularly when $\tan\beta$ is large. The constraint imposed by
measurements of $b\rightarrow s\gamma$ also excludes small
values of $m_{1/2}$. 
Finally, there are
regions of the $(m_{1/2}, m_0)$ plane that are favoured by
the BNL measurement \cite{newBNL} of $g_\mu - 2$ at the 2-$\sigma$ level, corresponding to 
a deviation  from the Standard Model 
calculation \cite{Davier} using $e^+ e^-$ data.  One should be  however 
aware that this constraint is still under active discussion.

The preferred range of the relic LSP density  has been altered
significantly by the recent improved determination of the allowable range
of the cold dark matter density obtained by combining WMAP and other
cosmological data:  $0.094 < \Omega_{CDM} < 0.129$ at the 2-$\sigma$
level \cite{WMAP}.  In the second panel of Fig.
\ref{fig:UHM}, we see the effect of imposing the WMAP range on the
neutralino density \cite{eoss,Baer,morewmap}.
We see immediately that (i) the cosmological regions are
generally much narrower, and (ii) the `bulk' regions at small $m_{1/2}$
and $m_0$ have almost disappeared, in particular when the laboratory
constraints are imposed. Looking more closely at the coannihilation
regions, we see that (iii) they are significantly truncated as well as
becoming much narrower, since the reduced upper bound on $\Omega_\chi h^2$
moves the tip where $m_\chi = m_{\tilde \tau}$ to smaller $m_{1/2}$
so that the upper limit is now $m_{1/2} \la 950$ GeV or $m_\chi \la 400$ GeV. 

\begin{figure}[h]
\includegraphics[height=2.3in]{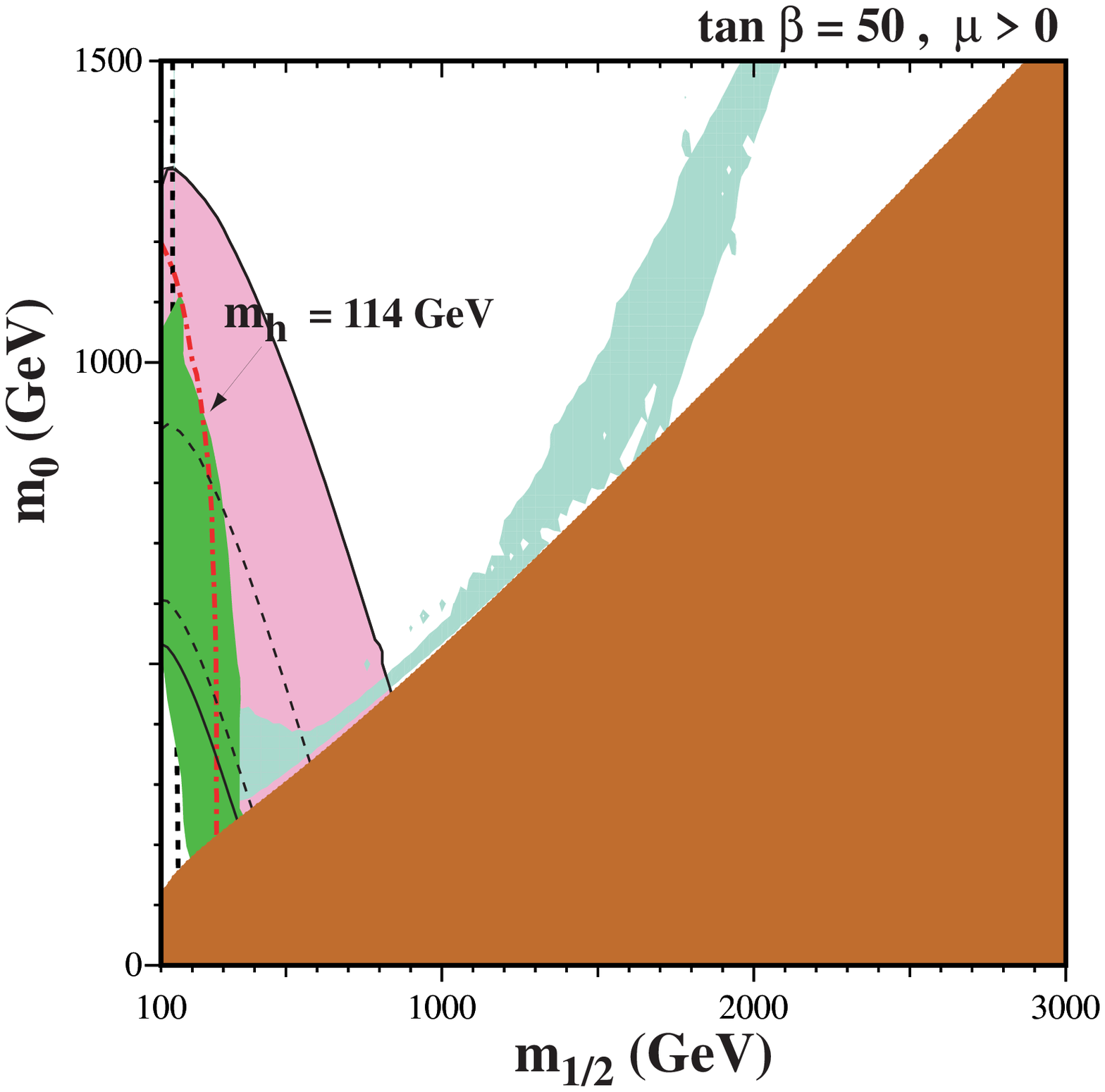}
\includegraphics[height=2.3in]{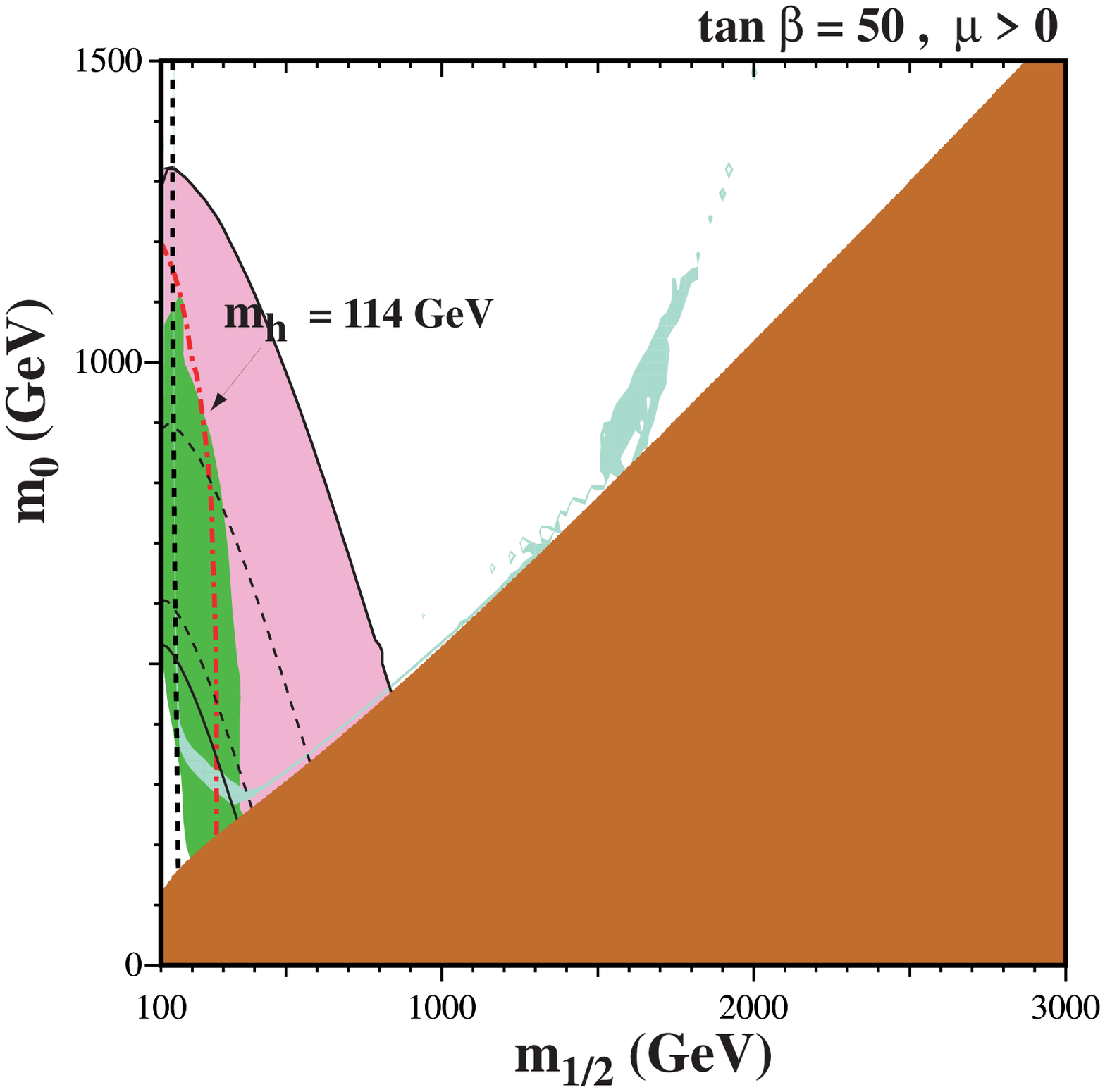}
\caption{\label{fig:UHM50}
{\it As in Fig. \protect{\ref{fig:UHM}} for $\tan \beta = 50$.}}
\end{figure}

Another
mechanism for extending the allowed CMSSM region to large
$m_\chi$ is rapid annihilation via a direct-channel pole when $m_\chi
\sim {1\over 2} m_{A}$~\cite{funnel,EFGOSi}. Since the heavy scalar and
pseudoscalar Higgs masses decrease as  
$\tan \beta$ increases, eventually  $ 2 m_\chi \simeq  m_A$ yielding a
`funnel' extending to large
$m_{1/2}$ and
$m_0$ at large
$\tan\beta$, as seen in the high $\tan \beta$ strips of Fig.~\ref{fig:UHM50}.
As one can see, the impact of the Higgs mass constraint is reduced (relative to 
the case with $\tan \beta = 10$) while that of $b \to s \gamma$ is enhanced.

\begin{figure}
\begin{center}
\includegraphics[height=3.2in]{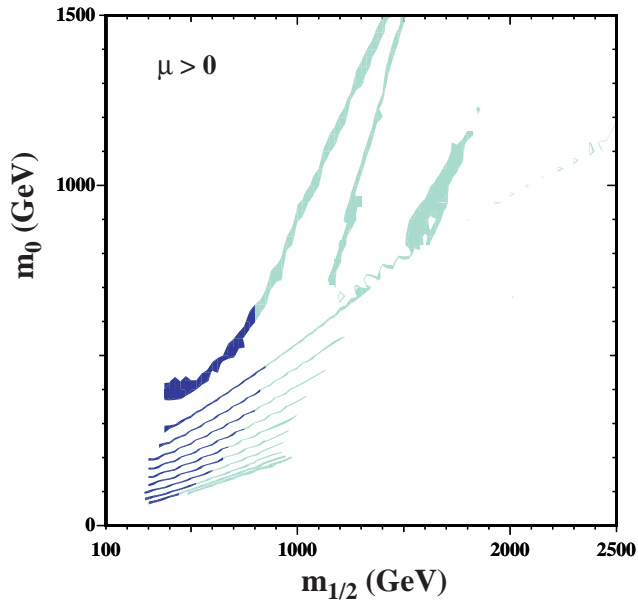}
\end{center}
\caption{\label{fig:strips}\it
The strips display the regions of the $(m_{1/2}, m_0)$ plane that are
compatible with $0.094 < \Omega_\chi h^2 < 0.129$ and the laboratory
constraints for $\mu > 0$ and $\tan \beta = 5, 10, 15, 20, 25, 30,
35, 40, 45, 50, 55$. The parts of the strips compatible with $g_\mu - 2$ 
at the 2-$\sigma$ level have darker shading.
}
\end{figure}

Shown in Fig.~\ref{fig:strips} are the WMAP lines \cite{eoss} of the $(m_{1/2}, m_0)$
plane allowed by the new cosmological constraint $0.094 < \Omega_\chi h^2
< 0.129$ and the laboratory constraints listed above, for $\mu > 0$ and
values of $\tan \beta$ from 5 to 55, in steps $\Delta ( \tan \beta ) = 5$.
We notice immediately that the strips are considerably narrower than the
spacing between them, though any intermediate point in the $(m_{1/2},
m_0)$ plane would be compatible with some intermediate value of $\tan
\beta$. The right (left) ends of the strips correspond to the maximal
(minimal) allowed values of $m_{1/2}$ and hence $m_\chi$. 
The lower bounds on $m_{1/2}$ are due to the Higgs 
mass constraint for $\tan \beta \le 23$, but are determined by the $b \to 
s \gamma$ constraint for higher values of $\tan \beta$.

\begin{figure}
\begin{center}
\includegraphics[height=2.6in]{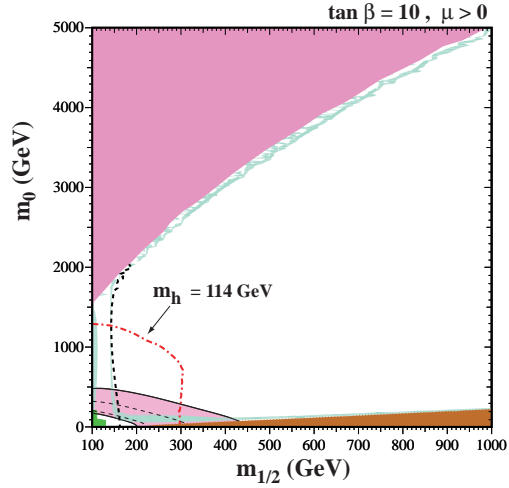}
\end{center}
\caption{\label{figfp}\it
As in Fig. \protect\ref{fig:UHM}a, where the range in $m_0$ is extended to
5 TeV.  In the shaded region at very high $m_0$, there are no solutions 
for $\mu$ which respect the low energy electroweak symmetry breaking conditions.
}
\end{figure}

\begin{figure}
\begin{center}
\includegraphics[height=2.6in]{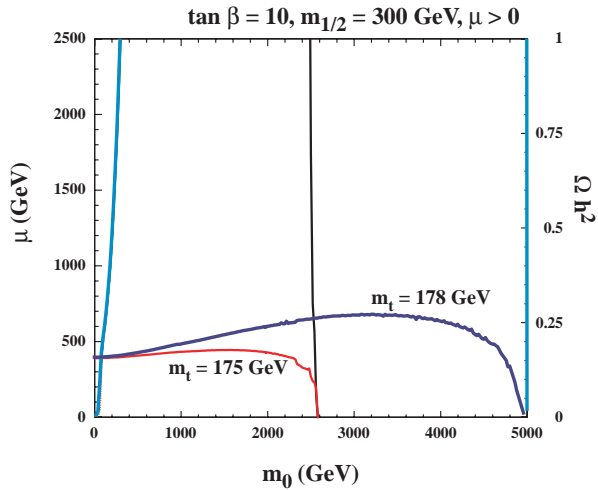}
\end{center}
\caption{\label{fignofp}\it
The value of $\mu$ as a function of $m_0$ for fixed $m_{1/2} = 300$ GeV
and $\tan \beta = 10$ for two choices of $m_t$ as indicated. The scale on the right
gives the value of $\Omega h^2$.  The curves corresponding to this is scale
rise sharply at low $m_0$ to values much larger than 1.  For $m_t = 175$ GeV and 
$m_0 \approx 2500$ GeV, the value of $\Omega h^2$ drops to acceptable values 
when $\mu$ becomes small.  When the $m_t = 178$ GeV,  
$\Omega h^2$ drops at $m_0 \approx 5000$ GeV.}
\end{figure}

Finally, there is one additional region of acceptable relic density known as the
focus-point region \cite{fp}, which is found
at very high values of $m_0$. An example showing this region is found in Fig. \ref{figfp},
plotted for $\tan \beta = 10$, $\mu > 0$, and $m_t = 175$ TeV.
As $m_0$ is increased, the solution for $\mu$ at low energies as determined
by the electroweak symmetry breaking conditions eventually begins to drop. 
When $\mu \la m_{1/2}$, the composition of the LSP gains a strong Higgsino
component and as such the relic density begins to drop precipitously.  These effects
are both shown in  Fig. \ref{fignofp} where the value of $\mu$ and $\Omega h^2$ 
are plotted as a function of $m_0$ for fixed $m_{1/2} = 300$ GeV and $\tan \beta = 10$. 
As $m_0$ is increased further, there are no longer any solutions for $\mu$.  This 
occurs in the shaded region in the upper left corner of Fig. \ref{figfp}. 

Fig. \ref{fignofp} also exemplifies the degree of fine tuning associated with the
focus-point region.  While the position of the focus-point region in the $m_0, m_{1/2}$
plane is not overly sensitive to supersymmetric parameters, it is highly sensitive to 
the top quark Yukawa coupling which contributes to the evolution of $\mu$ \cite{rs,ftuning}. 
As one can see in the figure, a change in $m_t$ of 3 GeV produces a shift
of about 2.5 TeV in $m_0$.  Note that the position of the focus-point region
is also highly sensitive to the value of $A_0/m_0$. In Fig. \ref{fignofp}, $A_0 = 0$
was chosen.  For $A_0/m_0 = 0.5$, the focus point shifts from 2.5 to 4.5 TeV
and moves to larger $m_0$ as $A_0/m_0$ is increased.

\section{A Likelihood analysis of the CMSSM}

Up to now, in displaying acceptable regions of cosmological density
in the $m_0, m_{1/2}$ plane,  it has been assumed that the input
parameters are known with perfect accuracy so that the relic density can
be calculated precisely.  While all of the beyond the standard model
parameters are completely unknown and therefore carry no
formal uncertainties, standard model parameters such as 
the top and bottom Yukawa couplings are known but do 
carry significant uncertainties.  Indeed, we saw that in the case of the 
focus-point region, there is an intense sensitivity 
of the relic density to the top quark Yukawa.  Other regions in the $m_0, m_{1/2}$
plane, such as those corresponding to the rapid annihilation funnels
are also very sensitive to the 3rd generation Yukawas.

The optimal way to combine the various constraints (both phenomenological and cosmological)
is via a likelihood
analysis, as has been done by some authors both before~\cite{DeBoer} and
after~\cite{Baer} the WMAP data was released. When performing such an analysis,
in addition to the formal experimental errors, it is also essential to
take into account theoretical errors, which introduce systematic
uncertainties that are frequently non-negligible. 
Recently, we have preformed an extensive likelihood analysis of the CMSSM \cite{eoss4}.
Included is the full likelihood function for the LEP Higgs search, as
released by the LEP Higgs Working Group. This includes the small
enhancement in the likelihood just beyond the formal limit due to the LEP
Higgs signal reported late in 2000. This was re-evaluated most recently
in~\cite{LEPHiggs}, and cannot be regarded as significant evidence for a
light Higgs boson. We have also taken into account the indirect information 
on $m_h$ provided by a global fit to the precision electroweak data. 
The likelihood function from this indirect source does not vary 
rapidly over the range of Higgs masses found in the CMSSM, but we 
included this contribution with the aim of completeness.

The interpretation of the combined Higgs likelihood, ${\cal L}_{exp}$,  
in the $(m_{1/2},
m_0)$ plane depends on uncertainties in the theoretical calculation of
$m_h$. These include the experimental error in $m_t$ and (particularly at
large $\tan \beta$) $m_b$, and theoretical uncertainties associated with
higher-order corrections to $m_h$. Our default assumptions are that $m_t =
175 \pm 5$~GeV for the pole mass, and $m_b = 4.25 \pm 0.25$~GeV for the
running $\overline {MS}$ mass evaluated at $m_b$ itself.
The theoretical uncertainty in $m_h$,  $\sigma_{th}$,  is dominated by
the experimental uncertainties in $m_{t,b}$, which are 
treated as uncorrelated Gaussian errors:
\beq
\sigma_{th}^2 = \left( \frac{\partial m_h}{\partial m_t} \right)^2 \Delta 
m_t^2 + \left( \frac{\partial m_h}{\partial m_b} \right)^2 \Delta m_b^2 \,.
\label{eq:sigmath}
\eeq
Typically, we find that $(\partial m_h/\partial m_t) \sim 0.5$, so that 
$\sigma_{th}$ is roughly 2-3 GeV.

The
combined experimental likelihood, ${\cal L}_{exp}$, from 
direct searches at LEP~2 and a global
electroweak fit is then convolved with a theoretical likelihood
(taken as a Gaussian) with uncertainty given by $\sigma_{th}$ from
(\ref{eq:sigmath}) above. Thus, we define the
total Higgs likelihood function, ${\cal L}_h$, as
\beq
{\cal L}_h(m_h) = { {\cal N} \over {\sqrt{2 \pi}\, \sigma_{th}  }}
 \int d m^{\prime}_h \,\, {\cal L}_{exp}(m^{\prime}_h)
 \,\, e^{-(m^{\prime}_h-m_h)^2/2 \sigma_{th}^2 }\, ,
\label{eq:higlik}
\eeq
where ${\cal N}$ is a factor that normalizes  the experimental likelihood
distribution. 

In addition to the Higgs likelihood function, we  have included the likelihood 
function based on $b \to s \gamma$. 
The branching ratio for these decays has been measured
by the CLEO, BELLE and BaBar collaborations~\cite{bsgex}, and we took as
the combined value ${\cal{B}}(b \to s \gamma)=(3.54 \pm 0.41 \pm
0.26)\times 10^{-4}$. The theoretical prediction 
\cite{gam,bsgth} contains uncertainties which stem from the uncertainties
in $m_b$, $\alpha_s$, the measurement of the semileptonic branching ratio
of the $B$ meson as well as the effect of the scale dependence.
While the likelihood function based on the measurements of the 
anomalous magnetic moment of the muon was considered in \cite{eoss4}, 
it will not be discussed here. 

Finally, in calculating the likelihood of the CDM density, we take into
account the contribution of the uncertainties in $m_{t,b}$. 
We will see that the theoretical uncertainty plays a
very significant role in this analysis. The likelihood for $\Omega h^2$ is therefore,
\beq
{\cal L}_{\Omega h^2}=\frac{1}{\sqrt{2 \pi} \sigma}
e^{-({\Omega h^2}^{th}-{\Omega h^2}^{exp})^2/2 \sigma^2} \,,
\label{eq:likamu}
\eeq 
where $\sigma^2=\sigma_{exp}^2+\sigma_{th}^2$, with
$\sigma_{exp}$ taken from the WMAP  \cite{WMAP}  result and $\sigma_{th}^2$
from (\ref{eq:sigmath}), replacing $m_h$ by $\Omega h^2$.

The total likelihood function is computed by combining all the components
described above: 
\beq {\cal L}_{tot} = {\cal L}_h \times {\cal
L}_{bs\gamma} \times {\cal L}_{\Omega_\chi h^2} (\times {\cal L}_{a_\mu})
\eeq
The likelihood function in the CMSSM can be considered  a function of
two variables, ${\cal L}_{tot}(m_{1/2},m_0)$, where $m_{1/2}$ and $m_0$
are the unified GUT-scale gaugino and scalar masses respectively.  
Results are based on a Bayesian analysis, in which a prior
range for $m_{1/2}$ is introduced in order to normalize the conditional probability
obtained from the likelihood function using  Bayes' theorem. Although it
is possible to motivate some upper limit on $m_{1/2}$, e.g., on the basis
of naturalness~\cite{nat,ftuning,eos2}, one cannot quantify any such limit
rigorously.  Within the selected range, we adopt a flat prior distribution
for $m_{1/2}$, and normalize the volume integral:
\beq
 \int  {\cal L}_{tot} \, dm_0 \, dm_{1/2} \; = \; 1
\eeq
for each value of $\tan \beta$, combining where appropriate both signs of
$\mu$.  We note that no such prior need be specified for $m_0$. For any
given value of $m_{1/2}$, the theory is well defined only up to some
maximum value of $m_0$, above which radiative electroweak symmetry
breaking is no longer possible. We always integrate up to that point,
adopting also a flat prior distribution for $m_0$.

\begin{figure}[h]
\includegraphics[height=1.9in]{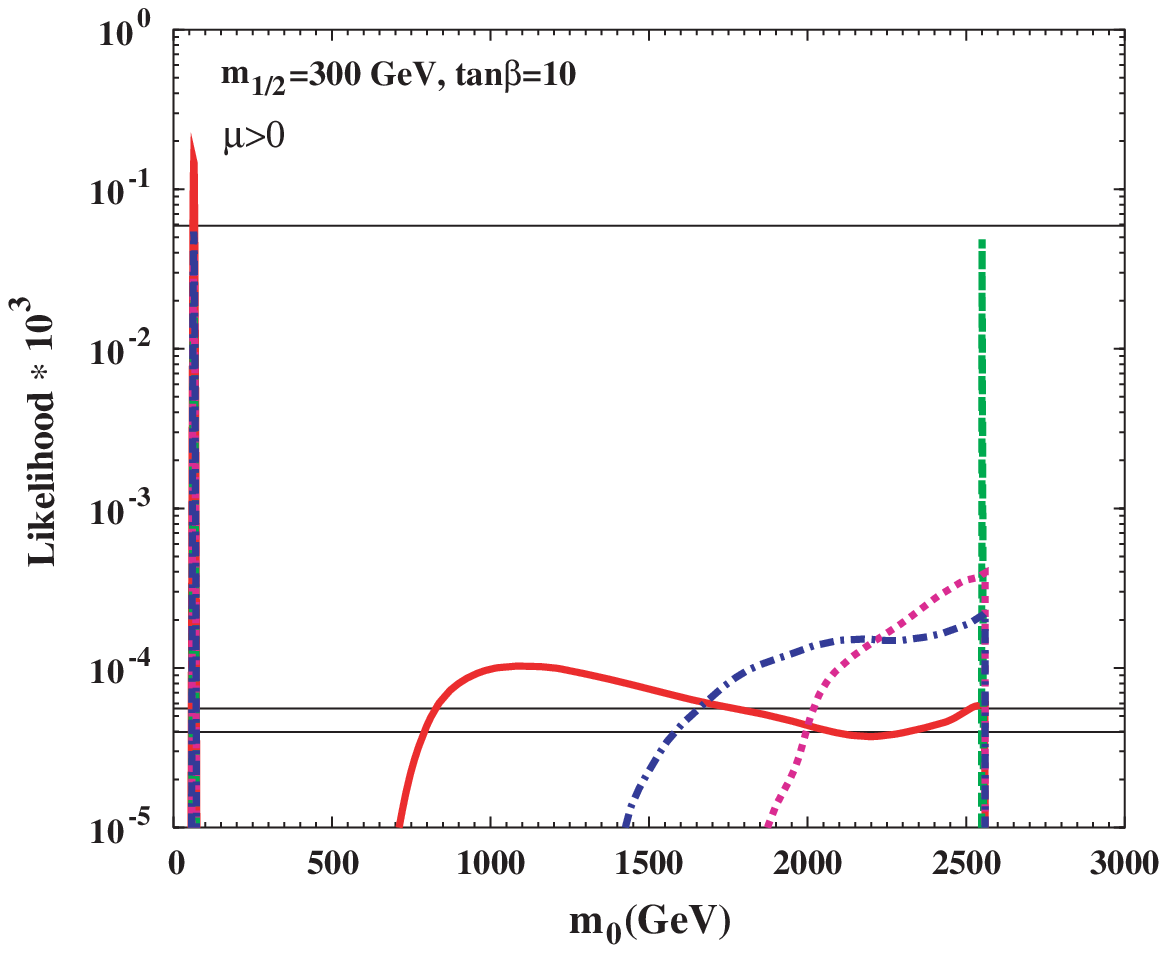}
\includegraphics[height=1.9in]{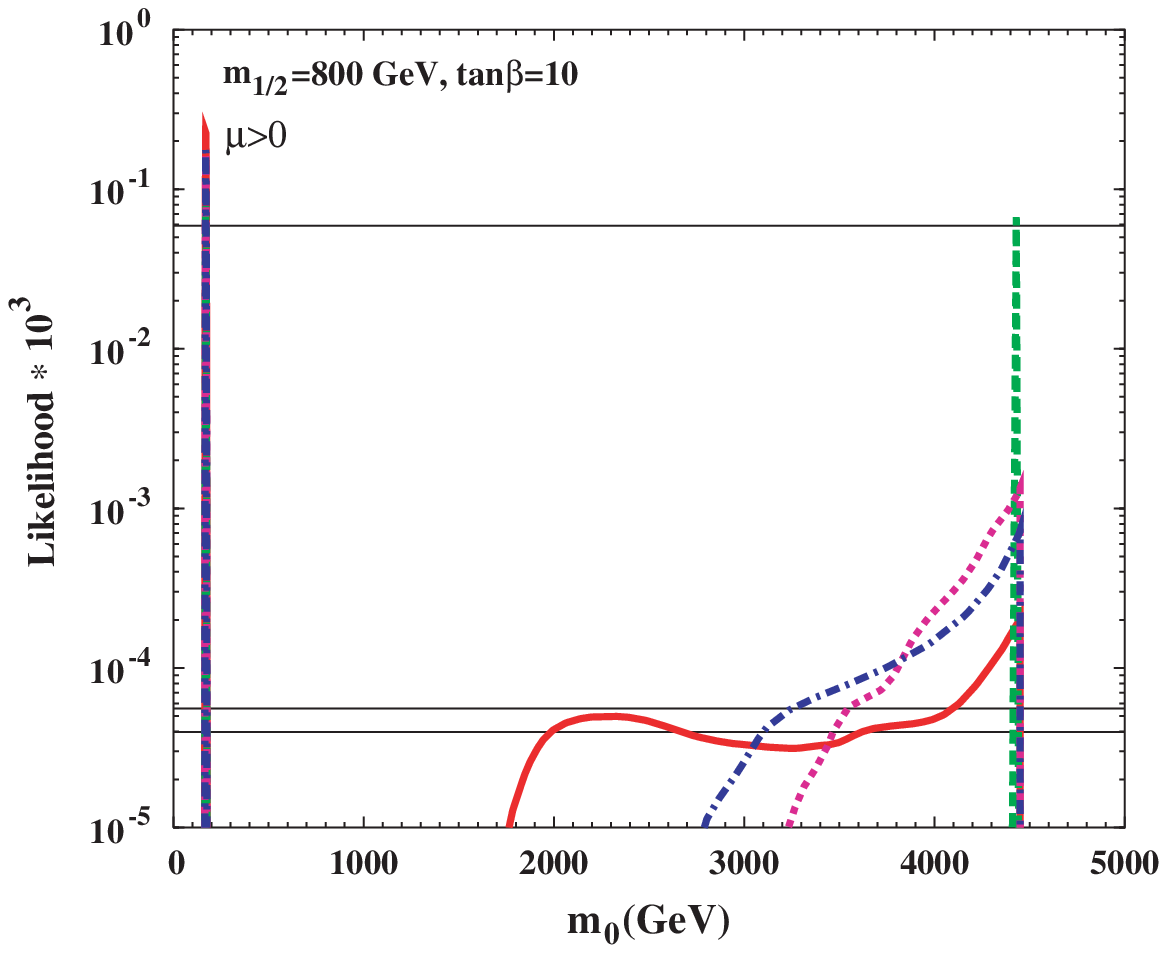}
\caption{\label{fig:WMAPFP}
{\it The likelihood function along slices in $m_0$ through the CMSSM parameter
space for $\tan \beta = 10, A_0 = 0, \mu> 0$ and $m_{1/2} = 300, 800$~GeV in the
left and right panels, respectively. The red (solid) curves are calculated using the current errors in 
$m_t$ and $m_b$, the green dashed curve with no error in $m_t$,
the violet dotted lines with $\Delta m_t = 0.5$~GeV, and
the blue dashed-dotted lines with $\Delta m_t = 1$~GeV. }}
\end{figure}

In  Fig.~\ref{fig:WMAPFP} the likelihood along slices
through the CMSSM parameter space for $\tan \beta = 10, A_0 = 0, \mu > 0,$
and $m_{1/2} = 300$ and 800~GeV is shown in the left and right panels,
respectively, plotting the likelihood as a function of $m_0$. 
The solid red curves show the
total likelihood function calculated including the uncertainties which
stem from the experimental errors in $m_t$ and $m_b$.
The peak at low $m_0$ is due to the
coannihilation region. The peak at $m_0 \simeq 2500 (4500)$ GeV for
$m_{1/2} = 300 (800)$ GeV is due to the focus-point region.
Also shown in Fig. \ref{fig:WMAPFP} are the 68\%, 90\%, and 95\% CL (horizontal)
lines, corresponding to the iso-likelihood values of the fully integrated
likelihood function corresponding to the solid (red) curve. 

The focus-point peak is suppressed relative to the coannihilation peak at
low $m_0$ because of the theoretical sensitivity to the experimental
uncertainty in the top mass.  We recall that the likelihood function is
proportional to $\sigma^{-1}$, and that $\sigma$ which scales with
$\partial (\Omega_\chi h^2 )/ \partial m_t$,  is very large at large
$m_0$~\cite{ftuning}. 
The error due
to the uncertainty in $m_t$ is far greater in the focus-point region than
in the coannihilation region.  Thus, even though the exponential in ${\cal
L}_{\Omega_\chi h^2}$ is of order unity near the focus-point region when
$\Omega_\chi h^2 \simeq 0.1$, the prefactor is very small due the large
uncertainty in the top mass. This accounts for the factor of $\ga 1000$ suppression
seen in Fig.~\ref{fig:WMAPFP} when comparing the two peaks of the solid red 
curves.

We note also that there is another broad, low-lying peak at intermediate
values of $m_0$. This is due to a combination of the effects of $\sigma$
in the prefactor and the exponential.  We expect a bump to occur when the
Gaussian exponential is of order unity, i.e., $\Omega_\chi h^2 \sim
\sqrt{2}\Delta m_t \, \partial \Omega_\chi h^2/\partial m_t$. 
$\Omega_\chi h^2 \sim 10$ at
large $m_0$ for our nominal value $m_t$ = 175 GeV, but it varies 
significantly as one samples the favoured range of $m_t$ within its 
present uncertainty.
The competition between the exponential and the prefactor
would require a large theoretical uncertainty in $\Omega_\chi h^2$: 
$\partial \Omega_\chi h^2/\partial m_t \sim 2$ for $\Delta m_t =
5$ GeV.  This
occurs when $m_0 \sim 1000$ GeV, which is the position of the broad
secondary peak in Fig.~\ref{fig:WMAPFP}a. At higher $m_0$, $\sigma$
continues to grow, and the prefactor suppresses the likelihood function
until $\Omega_\chi h^2$ drops to $\sim 0.1$ in the focus-point region.

As is clear from the above discussion, the impact of the present
experimental error in $m_t$ is particularly important in this region. This
point is further demonstrated by the differences between the curves in
each panel, where we decrease {\it ad hoc} the experimental uncertainty in
$m_t$.  As $\Delta m_t$ is decreased, the intermediate bump blends into
the broad focus-point peak.  
When the uncertainties in $m_t$ and $m_b$ are
set to 0, we obtain a narrow peak in the focus-point region. 

\begin{figure}[h]
\includegraphics[height=2.3in]{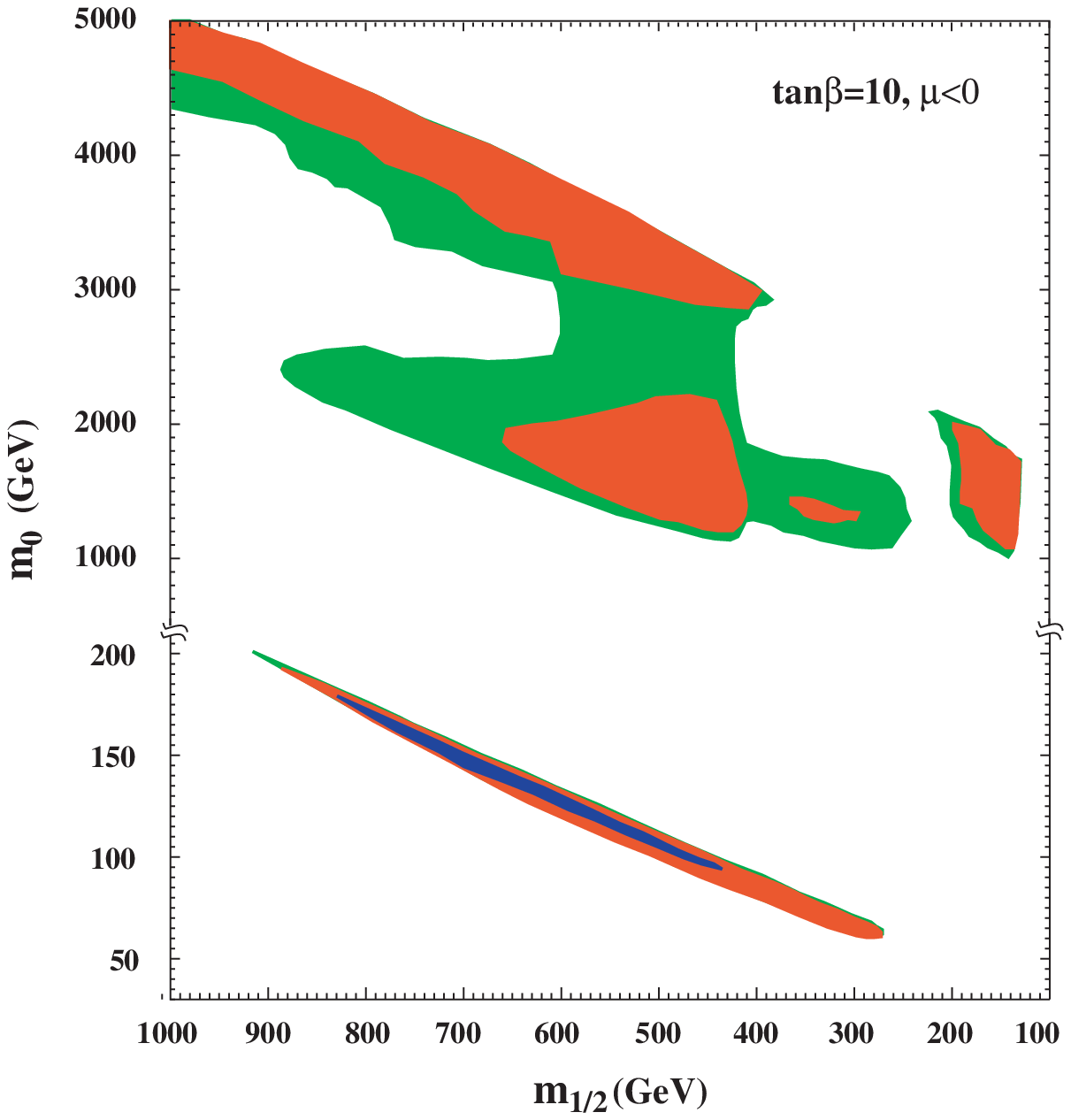}
\hspace {-.17in}
\includegraphics[height=2.25in]{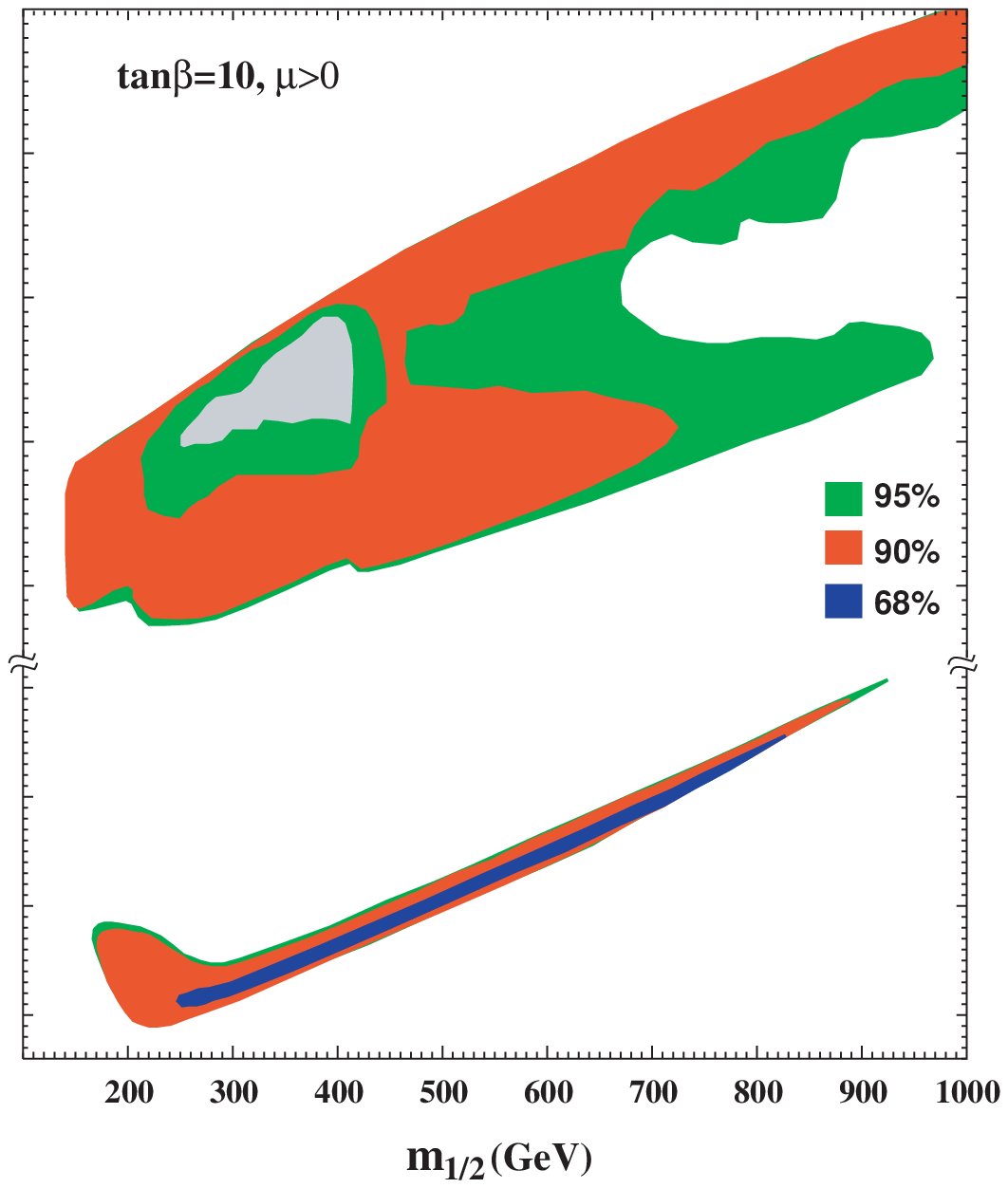}
\caption{\label{fig:contours}
{\it Contours of the likelihood at the 68\%, 90\% and 95\% levels for $\tan 
\beta = 10$, $A_0 = 0$ and $\mu > 0$ (left panel) or $\mu < 0$ (right 
panel), calculated 
using information of $m_h$, $b \to s \gamma$ and $\Omega_{CDM} h^2$ and 
the current uncertainties in $m_t$ and $m_b$. }}
\end{figure}

Using the fully normalized likelihood function ${\cal L}_{tot}$ obtained
by combining both signs of $\mu$ for each value of $\tan \beta$, we can
determine the regions in the $(m_{1/2}, m_0)$ planes which correspond to
specific CLs.
Fig.~\ref{fig:contours} extends the previous analysis to the entire
$(m_{1/2}, m_0)$ plane for $\tan \beta = 10$ and $A_0 = 0$, including both
signs of $\mu$. The darkest (blue), intermediate (red) and lightest
(green) shaded regions are, respectively, those where the likelihood is
above 68\%, above 90\%, and  above 95\%. 
Overall, the likelihood for $\mu < 0$ is less than that for 
$\mu > 0$ due to the Higgs and $b \to s \gamma$ constraints. 
Only the bulk and coannihilation-tail regions appear above the 68\% level, 
but the focus-point region appears above the 90\% level, and so cannot be 
excluded.

\begin{figure}[h]
\includegraphics[height=2.3in]{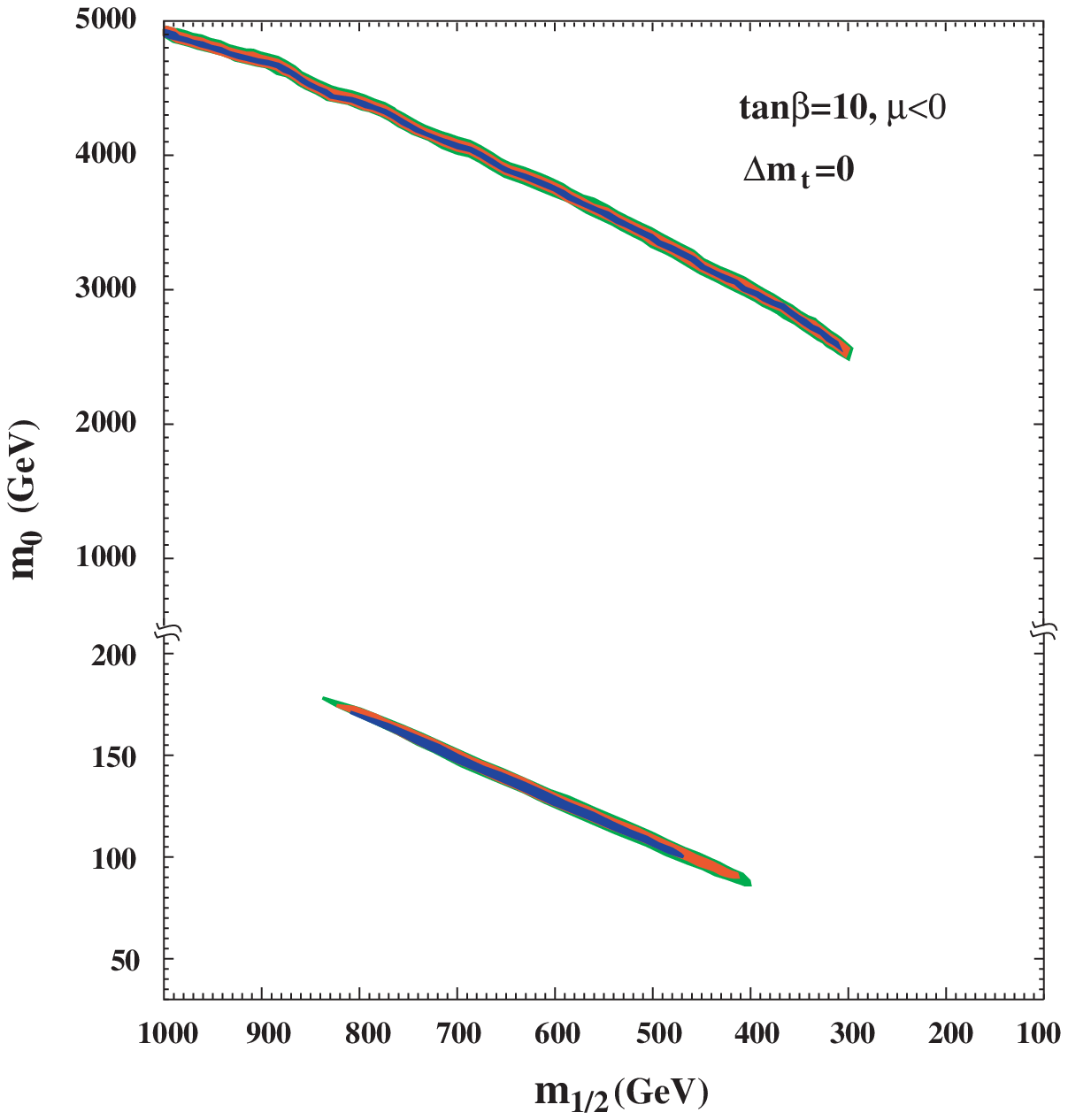}
\hspace {-.17in}
\includegraphics[height=2.25in]{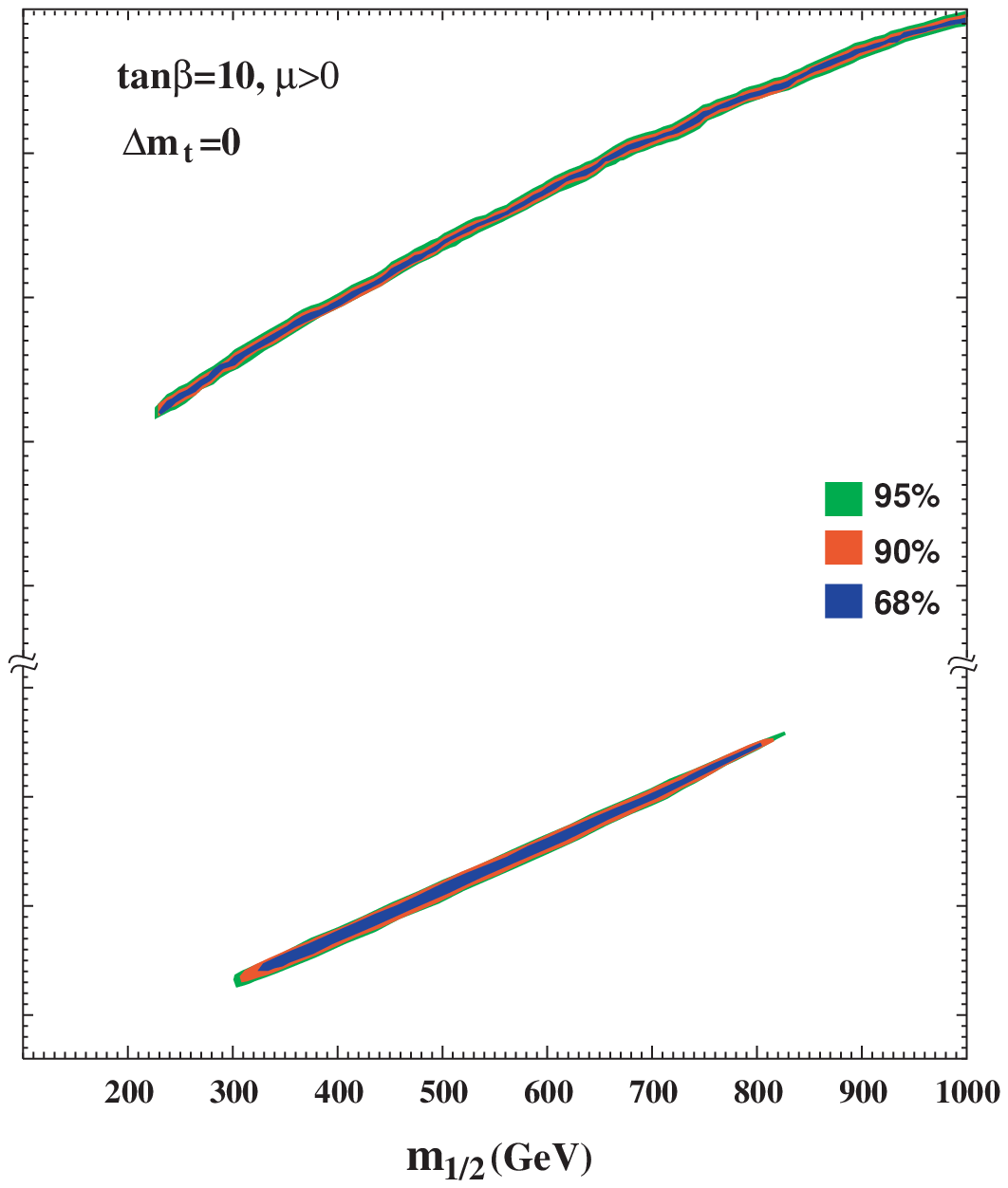}
\caption{\label{fig:contourswithoutmt}
{\it As in Fig.~\protect\ref{fig:contours} but assuming zero uncertainty in 
$m_t$.}}
\end{figure}

The bulk region is more apparent in the right panel of
Fig.~\ref{fig:contours} for $\mu > 0$ than it would be if the experimental
error in $m_t$ and the theoretical error in $m_h$ were neglected. 
Fig.~\ref{fig:contourswithoutmt} complements the previous figures by 
showing the likelihood functions as they would appear if there were no 
uncertainty in $m_t$, keeping the other inputs the same. 
We see that, in this case, both the 
coannihilation and focus-point strips rise above the 68\% CL.

\begin{figure}[h]
\centering
\includegraphics[height=2.3in]{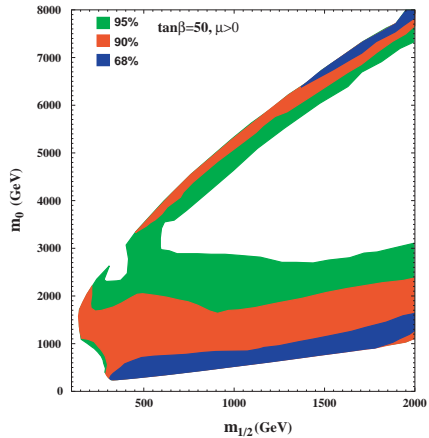}
\caption{\label{fig:contours50}
{\it Likelihood contours as in Fig.~\ref{fig:contours}, but for $\tan 
\beta = 50$, $A_0 = 0$ and $\mu> 0$.}}
\end{figure}

Fig.~\ref{fig:contours50} shows the likelihood projection for $\tan \beta = 50$, 
$A_0 = 0$ and $\mu >0$.   In this case, 
regions at small $m_{1/2}$ and $m_0$ are disfavoured by the $b \to s
\gamma$ constraint.  The coannihilation
region is broadened by a merger with the rapid-annihilation funnel.
Both the coannihilation and
the focus-point regions feature strips allowed at the 68\% CL, and these
are linked by a bridge at the 95\% CL.

\section{Beyond the CMSSM}

\begin{figure}[hbtp]
\includegraphics[height=2.3in]{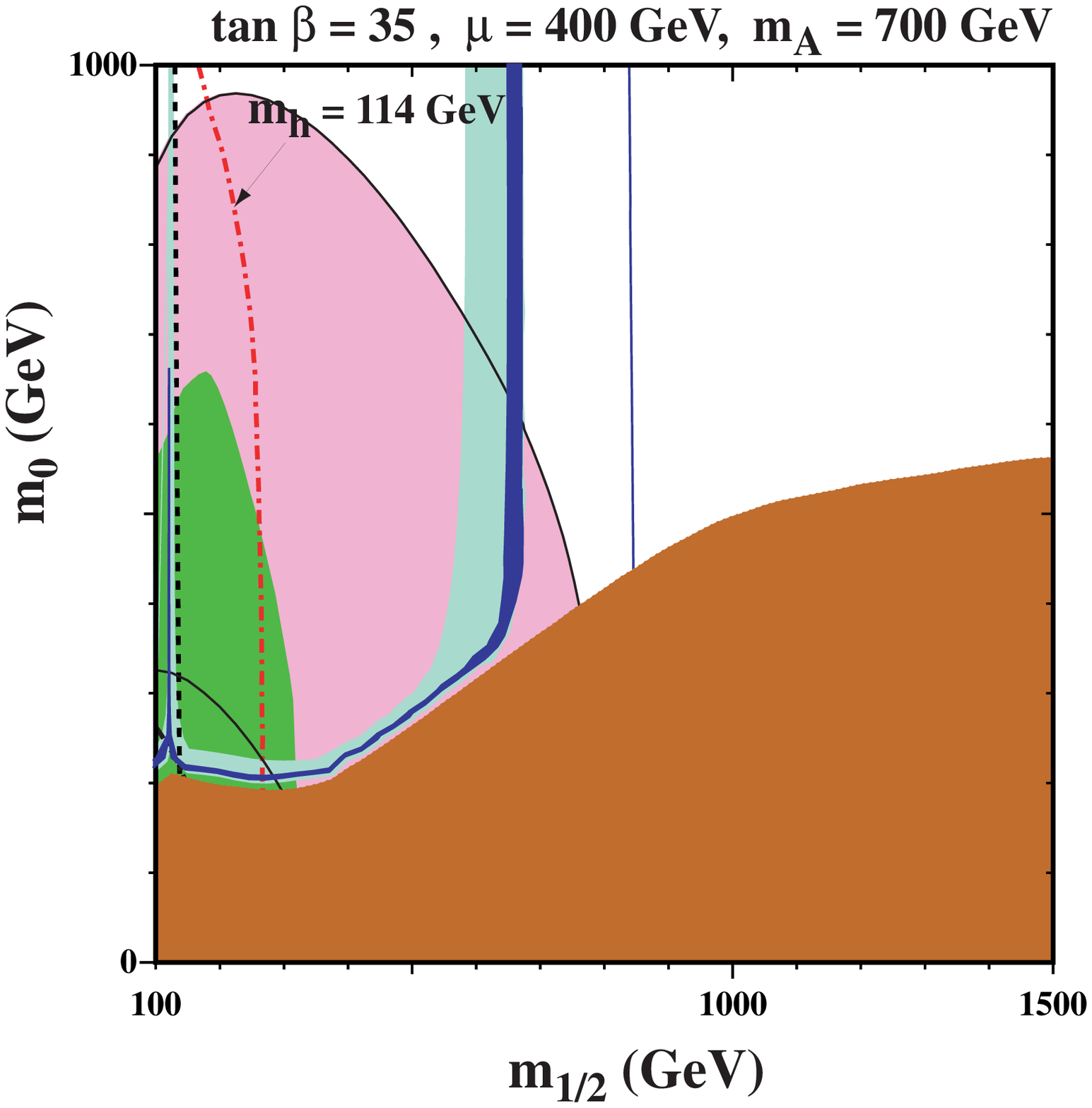}
\includegraphics[height=2.3in]{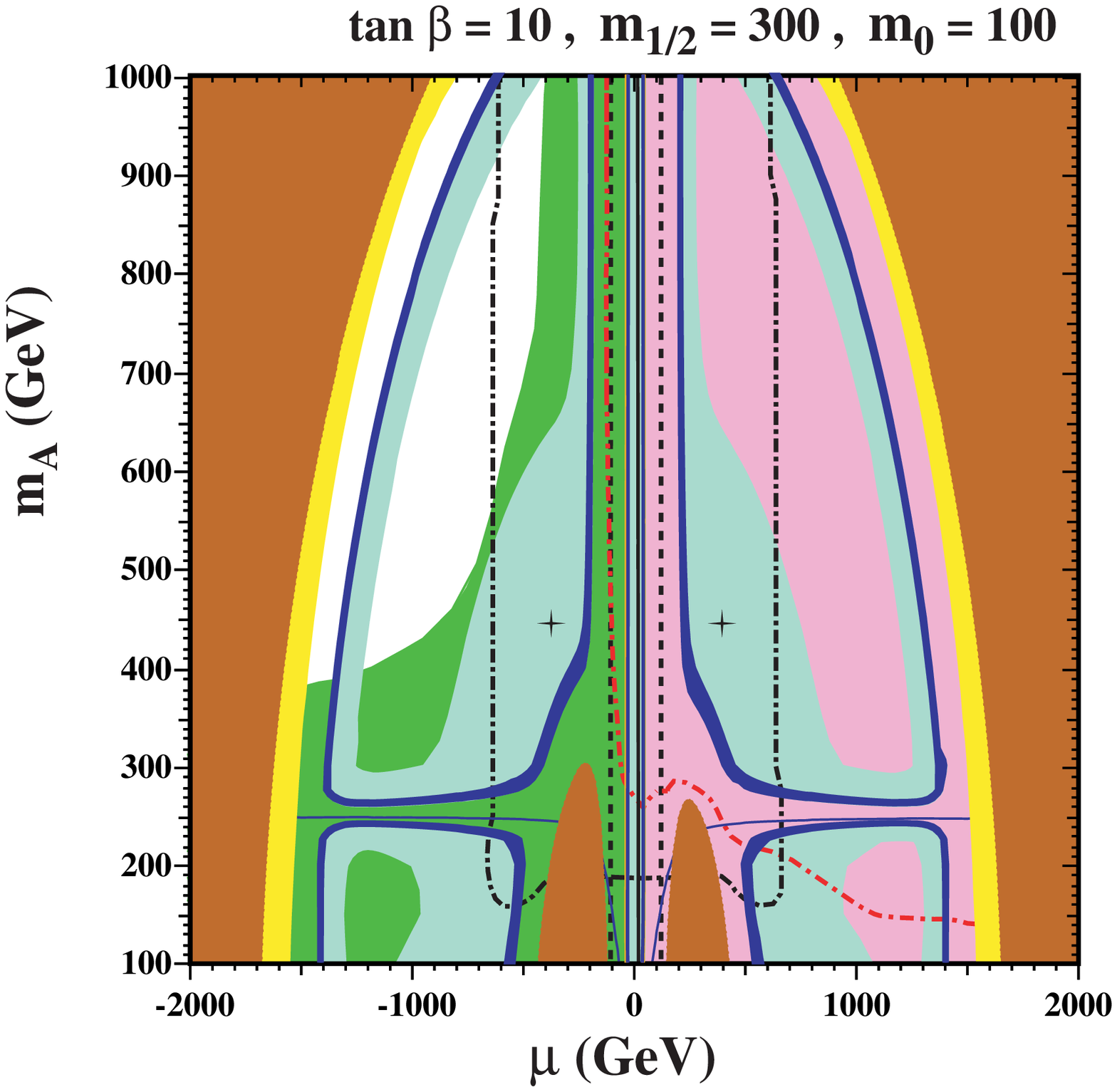}
\caption{\it   a) The NUHM $(m_{1/2}, m_0)$ plane for $\tan \beta = 35$, (a) $\mu = 400$~GeV 
and $m_{A} = 700$~GeV  b)the NUHM $(\mu, m_A)$ plane for $\tan \beta = 10$, $m_0
= 100$~GeV and $m_{1/2} = 300$~GeV,  with $A_0 = 0$.
The (red)
dot-dashed lines are the contours $m_h = 114$~GeV, and the near-vertical
(black) dashed lines are the contours $m_{\chi^\pm} = 103.5$~GeV. The
dark (black) dot-dashed lines indicate the GUT stability constraint. Only
the areas inside these curves (small $\mu$) are allowed by this
constraint. The light (turquoise) shaded areas are the cosmologically
preferred regions with
\protect\mbox{$0.1\leq\ohsq\leq 0.3$}. 
The darker (blue) portion of this region corresponds to the newer
WMAP densities.  The dark (brick red) shaded
regions is excluded because a charged particle is lighter than the 
neutralino, and the lighter (yellow) shaded regions is excluded because
the LSP is a sneutrino. The medium
(green) shaded region is excluded by $b \to s \gamma$.
The regions allowed by the $g-2$ constraint 
are shaded (pink) and bounded by solid black lines. The solid (blue)
curves correspond to $m_\chi = m_A/2$. }
	\label{muma}
\end{figure}

The results of  the CMSSM described in the previous sections are based heavily
on the assumptions of universality of the supersymmetry breaking parameters.
One of the simplest generalizations of this model relaxes the assumption of
universality of the Higgs soft masses and is known as the NUHM \cite{eos3}
In this case, the 			
input parameters include $ \mu$ and $m_A,$ in addition to the standard CMSSM inputs.
In order to switch $\mu$ and $m_A$ from outputs to inputs,
the two soft Higgs masses,  $m_1,
m_2$ can no longer be set equal to $m_0$ and instead are 
calculated from the electroweak symmetry breaking conditions. The NUHM parameter space was recently analyzed \cite{eos3} and a sample of the results
are shown in Fig. \ref{muma}.

In the left panel of Fig. \ref{muma}, we see a $m_{1/2},m_0$ plane
with a relative low value of $\mu$.  In this case, an allowed region is found
when the LSP contains a non-negligible Higgsino component
which moderates the relic density independent of $m_0$.  To the right of this region,
the relic density is too small.  In the right panel, we see an example of the $m_A,\mu$
plane.  The crosses correspond to CMSSM points.  In this single pane, we
see examples of acceptable cosmological regions corresponding to the 
bulk region, co-annihilation region and s-channel annihilation through the Higgs
pseudo scalar.

Rather than relax the CMSSM, it is in fact possible to further constrain the model.
While the CMSSM models described above are certainly
mSUGRA inspired, minimal supergravity models
can be argued to be still more predictive.  
Let us assume that supersymmetry is broken in a hidden sector
so that the superpotential can be written as a sum of two terms, $W = F(\phi) +g(\zeta)$,
where $\phi$ represents all observable fields and $\zeta$ all hidden sector fields.
We furthermore must choose $g(\zeta)$ such that when $\zeta$ picks up a vacuum
expectation value, supersymmetry is broken.
When the potential is expanded and terms inversely proportional to Planck mass are dropped,
one finds \cite{BIM} 1) scalar mass universality with $m_0 = \langle g \rangle$; 
2) trilinear mass universality with 
$A_0 = \langle dg/d\zeta \rangle \langle \zeta \rangle + \langle g \rangle \langle \zeta \rangle^2$; 
and 3) $B_0 = A_0 - m_0$.

In the simplest version of the theory \cite{pol}, the
universal trilinear soft
supersymmetry-breaking terms  are $A = (3 -
\sqrt{3}) m_{0}$ and bilinear
soft supersymmetry-breaking term is $B = (2 - \sqrt{3}) m_{0}$, i.e., a
special case of the general relation  above between $B$ and
$A$. 

Given a relation between $B_0$ and $A_0$, we can no longer use the
standard CMSSM boundary conditions, in which $m_{1/2}$, $m_0$, $A_0$,
$\tan \beta$, and $sgn(\mu)$ are input at the GUT scale with $\mu$ and $B$ 
determined by the electroweak symmetry breaking condition.
Now, one is forced to input $B_0$ and instead $\tan \beta$ is 
calculated from the minimization of the Higgs potential \cite{eoss2}.

\begin{figure}
\includegraphics[height=2.3in]{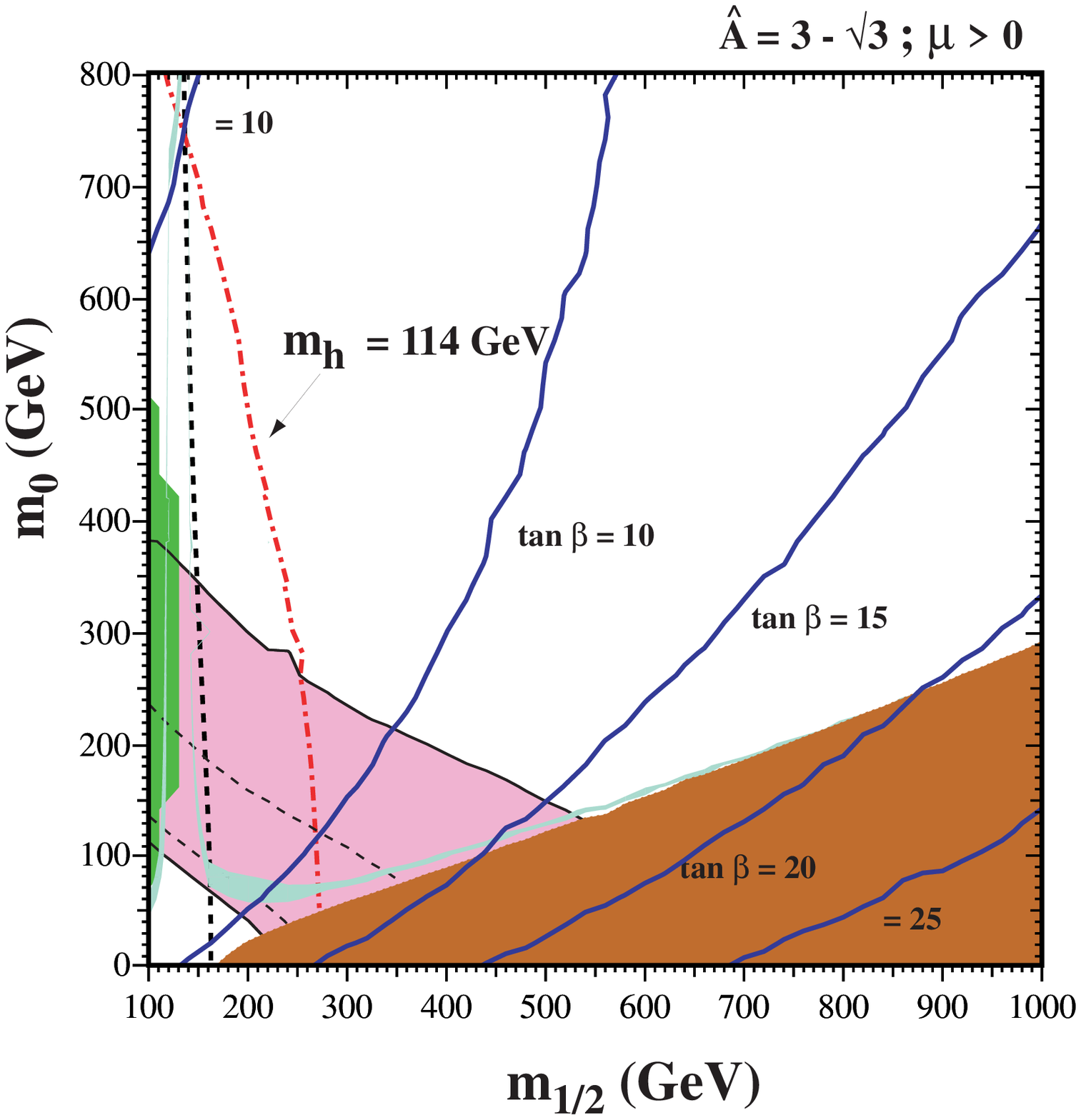}
\includegraphics[height=2.3in]{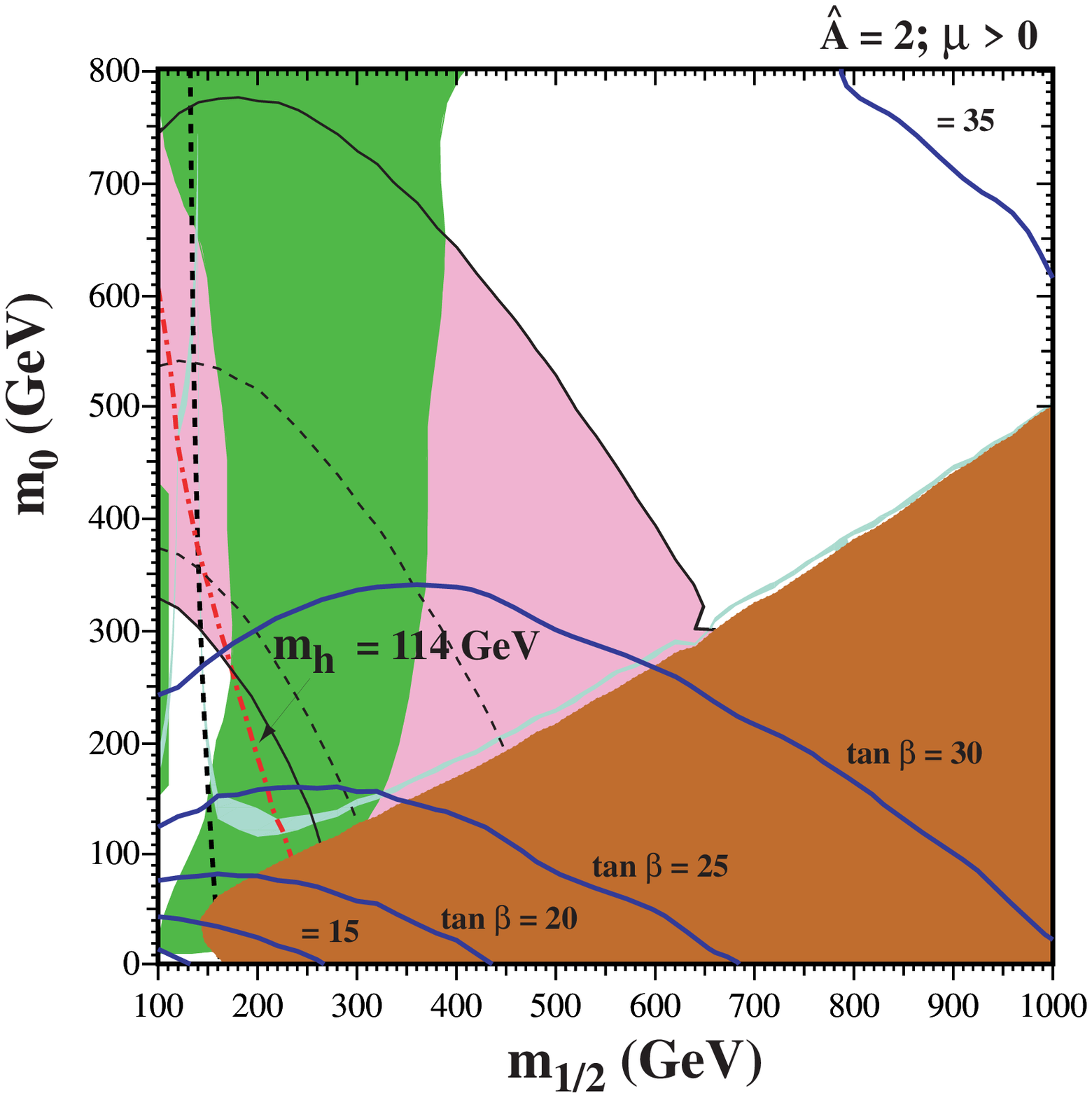}
\caption{\it
Examples of $(m_{1/2}, m_0)$ planes with contours of $\tan \beta$ 
superposed, for $\mu > 0$ and (a)  the simplest Polonyi model with ${\hat A} = 3 - 
\sqrt{3}, {\hat B} = {\hat A} -1$ and (b) ${\hat A} = 2.0, {\hat B} =
{\hat A} -1$. In each panel, we show the regions excluded by 
the LEP lower limits on MSSM particles, those ruled out by $b
\to s \gamma$ decay (medium green shading), and those 
excluded 
because the LSP would be charged (dark red shading). The region favoured 
by the WMAP range  has light turquoise shading. The region 
suggested by $g_\mu - 2$ is medium (pink) shaded.}
\label{fig:Polonyi}
\end{figure}

In Fig.~\ref{fig:Polonyi}, the contours of $\tan \beta$ (solid
blue lines) in the $(m_{1/2}, m_0)$ planes for two values of ${\hat
A}  = A_0/m_0$, ${\hat B} = B_0/m_0 = {\hat A} - 1$ and the sign of $\mu$ are displayed \cite{eoss2}. 
Also shown are the contours where
$m_{\chi^\pm} > 104$~GeV (near-vertical black dashed lines) and $m_h >
114$~GeV (diagonal red dash-dotted lines). The excluded regions where
$m_\chi > m_{\tilde \tau_1}$ have dark (red) shading, those excluded by $b
\to s \gamma$ have medium (green) shading, and those where the relic
density of neutralinos lies within the WMAP range $0.094 \le \Omega_\chi
h^2 \le 0.129$ have light (turquoise) shading. Finally, the regions
favoured by $g_\mu - 2$ at the 2-$\sigma$ level are medium (pink) shaded.

In panel (a) of Fig.~\ref{fig:Polonyi}, we
see that the Higgs constraint combined with the relic density requires $\tan \beta
\ga 11$, whilst the relic density also enforces $\tan \beta \la 
20$. For a given point in the $m_{1/2} - m_0$ plane, the calculated value of $\tan \beta$
increases as ${\hat A}$ increases.
This is seen in panel (b) of
Fig.~\ref{fig:Polonyi}, when ${\hat A} = 2.0$, close to its maximal value
for $\mu > 0$, the $\tan \beta$ contours turn over towards smaller
$m_{1/2}$, and only relatively large values $25 \la \tan \beta \la 35$ are
allowed by the $b \to s \gamma$ and $\Omega_{CDM} h^2$ constraints,
respectively.  For any given value of ${\hat A}$, there is only a relatively narrow
range allowed for $\tan \beta$.

\section{Detectability}

The question of detectability with respect to supersymmetric models is of
key importance particularly with the approaching start of the LHC.
As an aid to the assessment of the prospects for detecting sparticles at
different accelerators, benchmark sets of supersymmetric parameters have
often been found useful, since they provide a focus for
concentrated discussion \cite{oldbench,SPS,newbench}.
A set of proposed post-LEP benchmark
scenarios \cite{oldbench} were chosen to span the CMSSM.  Five of the
chosen points are in the
`bulk' region at small $m_{1/2}$ and $m_0$, four are spread along the
coannihilation `tail' at larger $m_{1/2}$ for various values of
$\tan\beta$.  
Two points are in rapid-annihilation `funnels' at large $m_{1/2}$ and $m_0$. 
 Two points
were chosen in the focus-point region at large $m_0$. The proposed
points range over the allowed values of
$\tan\beta$ between 5 and 50. 

\begin{figure}[h]
\centering
\includegraphics[height=3.2in]{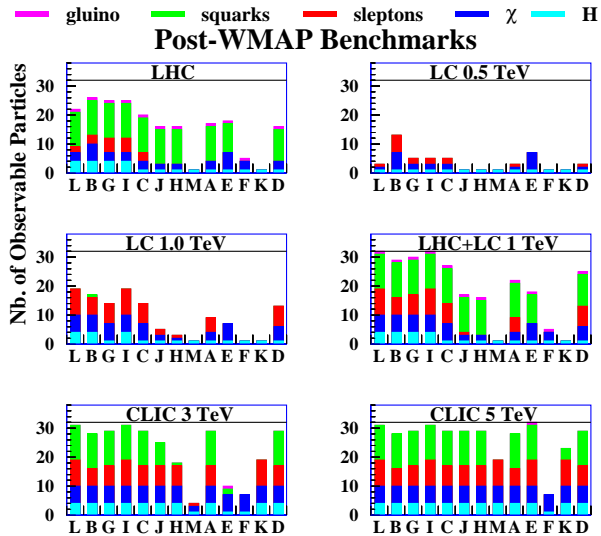}
\caption{\label{fig:newM}
{\it 
Summary of the numbers of MSSM particles that may be detectable at various
accelerators in the updated benchmark scenarios. 
We see that the capabilities of the LHC and of linear $e^+ e^-$ colliders 
are largely complementary. We re-emphasize that mass and coupling measurements
at $e^+ e^-$ colliders are usually much cleaner and more precise than at
hadron-hadron colliders such as the LHC, where, for example, it is not
known how to distinguish the light squark flavours. 
}}
\end{figure}

In Fig.~\ref{fig:newM}, a
comparison of the numbers of different MSSM particles that should
be observable at different accelerators in the various benchmark
scenarios~\cite{newbench}, ordered by their consistency with $g_\mu -2$. 
The qualities of the
prospective sparticle observations at hadron colliders and linear $e^+
e^-$ colliders are often very different, with the latters' clean
experimental environments providing prospects for measurements with better
precision.  Nevertheless, Fig.~\ref{fig:newM} already restates the clear
message that hadron colliders and linear $e^+ e^-$ colliders are largely
complementary in the classes of particles that they can see, with the
former offering good prospects for strongly-interacting sparticles such as
squarks and gluinos, and the latter excelling for weakly-interacting
sparticles such as charginos, neutralinos and sleptons. 

\begin{figure}[h]
\centering
\includegraphics[height=3.2in]{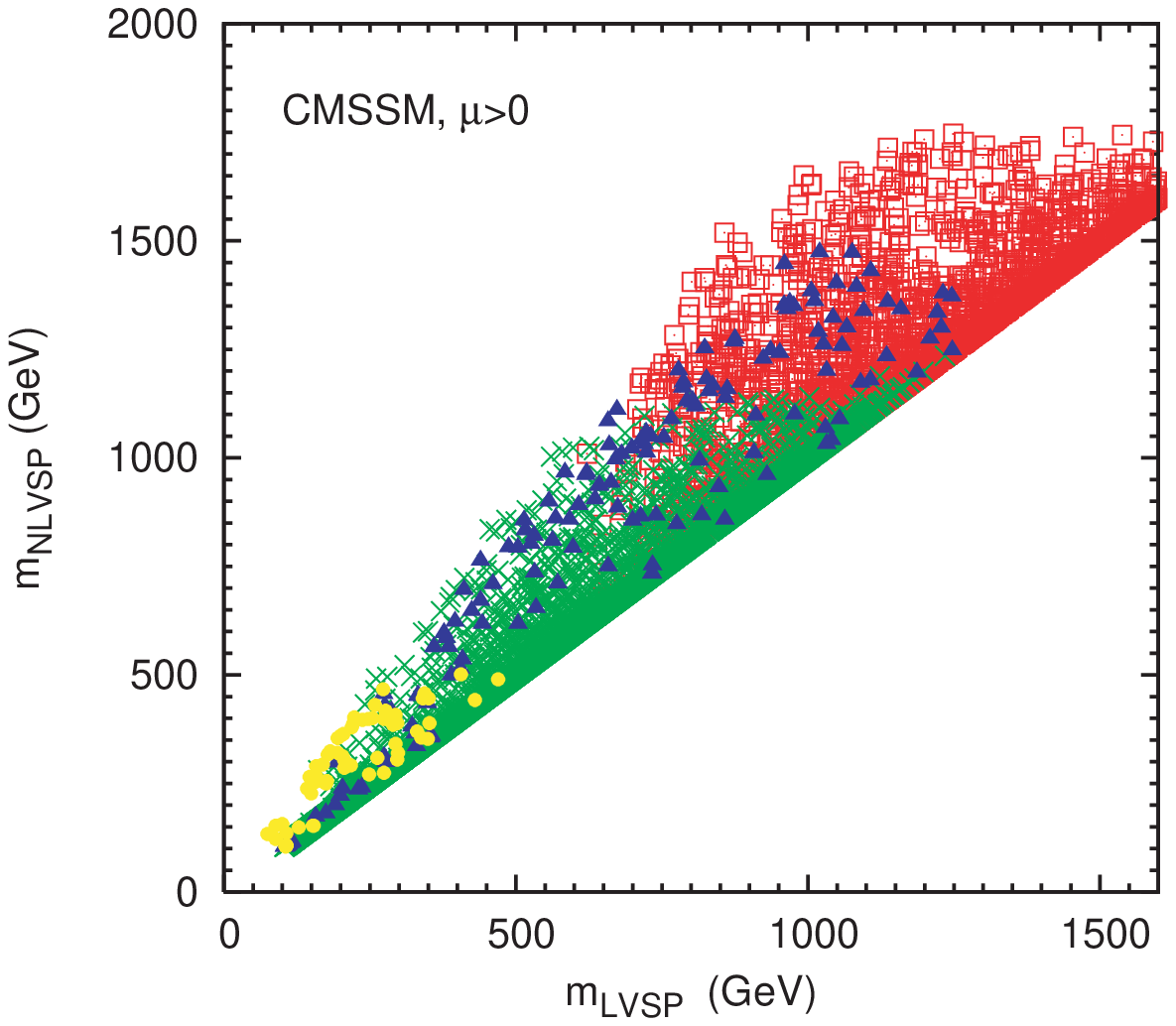}
\caption{
{\it 
Scatter plots of the masses of the lightest visible
supersymmetric particle (LVSP) and the next-to-lightest visible
supersymmetric particle (NLVSP) in the CMSSM for $\mu > 0$. 
The darker (blue) 
triangles satisfy all the laboratory, astrophysical and cosmological 
constraints. For comparison, the dark (red) squares and medium-shaded 
(green) crosses respect the laboratory 
constraints, but not those imposed by astrophysics and cosmology. 
In addition, the (green) crosses represent models which are expected to be
visible at the LHC. The 
very light (yellow) points are those for which direct detection of 
supersymmetric dark matter might be possible.} 
\label{fig:VSP10p} }
\end{figure}

Clearly the center of mass energy of any future linear collider is paramount
towards the supersymmetry discovery potential of the machine. 
This is seen in Fig. \ref{fig:newM} for the benchmark points as more sparticles
become observable at higher CM energy.  We can emphasize this point in general models
by plotting the masses of the two lightest (observable) sparticles in 
supersymmetric models.  For example, 
in Fig. \ref{fig:VSP10p} \cite{eoss7}, a scatter plot
of the masses of the lightest visible
supersymmetric particle (LVSP) and the next-to-lightest visible
supersymmetric particle (NLVSP) is shown for the CMSSM. 
Once again, points selected satisfy all phenomenological constraints.
We do not consider the LSP itself to be
visible, nor any heavier neutral sparticle that decays invisibly inside
the detector, such as ${\tilde \nu} \to \nu \chi$  when ${\tilde \nu}$ is the 
next-to-lightest sparticle in a 
neutralino LSP scenario. The LVSP and the NLVSP are the lightest sparticles likely to be
observable in collider experiments.

 All
points shown in Fig. \ref{fig:VSP10p} satisfy the phenomenological constraints discussed above.  
The dark (red) squares represent those points for which the relic density
is outside the WMAP range, and for which all coloured sparticles (squarks
and gluinos) are heavier than 2 TeV. The CMSSM parameter reach at the LHC
has been analyzed in~\cite{Baer2}. To
within a few percent accuracy, the CMSSM reach contours presented
in~\cite{Baer2} coincide with the 2-TeV contour for the lightest squark (generally the
stop) or gluino, so we regard the dark (red) points as unobservable at the
LHC. Most of these points have $m_{NLVSP} \ga 1.2$~TeV. Conversely, the
medium-shaded (green) crosses represent points where at least one squark
or gluino has a mass less than 2 TeV and should be observable at the LHC. 
The spread of the dark (red) squares and
medium-shaded (green) crosses, by as much as 500~GeV or more in some
cases, reflects the maximum mass splitting between the LVSP and the NLVSP
that is induced in the CMSSM via renormalization effects on the input mass
parameters. The amount of this spread also reflects our cutoff $|A_0| < 1$
TeV, which controls the mass splitting of the third generation sfermions.

The darker (blue) triangles are those points respecting the cosmological
cold dark matter constraint. Comparing with the regions populated by dark
(red)  squares and medium-shaded (green) crosses, one can see which of
these models would be detectable at the LHC, according to the criterion in
the previous paragraph. We see immediately that the dark matter constraint
restricts the LVSP masses to be less than about 1250~GeV and NLVSP masses
to be less than about 1500~GeV. In most cases, the identity of the LVSP
is the lighter $\tilde \tau$. While pair-production of the LVSP would
sometimes require a CM energy of about 2.5~TeV, in some cases there 
is a lower supersymmetric threshold due to
the associated production of the LSP $\chi$ with the 
next lightest 
neutralino $\chi_2$~\cite{djou}. Examining the masses and 
identities of the sparticle
spectrum at these points, we find that $E_{CM} \ga 2.2$ TeV would be 
sufficient to see 
at least one sparticle, as shown in Table~1. Similarly,  only a LC with 
$E_{CM} \ge
2.5$~TeV would be `guaranteed' to see two visible sparticles (in addition
to the $\chi$ LSP), somewhat lower than the 3.0 TeV one might
obtain by requiring the pair production of the NLVSP. 
Points with $m_{LVSP} \ga 700$ GeV are
predominantly due to rapid annihilation via direct-channel $H,A$ poles,
while points with 200 GeV $\la m_{LVSP} \la$ 700 GeV are largely due to
$\chi$-slepton coannihilation. 

\begin{table}[htb]
\begin{center}
\caption{\it Centre-of-mass energy (in TeV) required to observe 
one or two sparticles at a future LC in the CMSSM and NUHM.}
\label{tab:alpha}
\vskip .3cm
\begin{tabular}{|c|c|c|c|}
\hline {\it Model} &
$sgn(\mu)$  & {\it one sparticle} & {\it two sparticles} \\
\hline CMSSM & $\mu > 0 $ &  2.2  & 2.6  \\
&$\mu < 0$& 2.2 & 2.5 \\ \hline
 NUHM & $\mu > 0 $ &  2.4  & 2.8 \\
&$\mu < 0$& 2.6& 2.9 \\ \hline
\end{tabular}
\end{center}
\end{table}

An $E_{CM} = 500$~GeV LC would be able to explore the `bulk' region at low
$(m_{1/2}, m_0) $, which is represented by the small cluster of points
around $m_{LVSP} \sim 200$ GeV. It should also be noted that there are a
few points with $m_{LVSP} \sim 100$ GeV which are due to rapid
annihilation via the light Higgs pole. These points all have very large
values of $m_0$ which relaxes the Higgs mass and chargino mass
constraints, particularly when $m_t = 178$ GeV. A LC with $E_{CM} =
1000$~GeV would be able to reach some way into the coannihilation `tail',
but would not cover all the WMAP-compatible dark (blue) triangles. Indeed,
about a third of these points are even beyond the reach of the LHC in this
model. Finally, the light (yellow) filled circles are points for which the
elastic $\chi$-$p$ scattering cross section is larger than $10^{-8}$~pb.

\begin{figure}[h]
\includegraphics[height=1.65in]{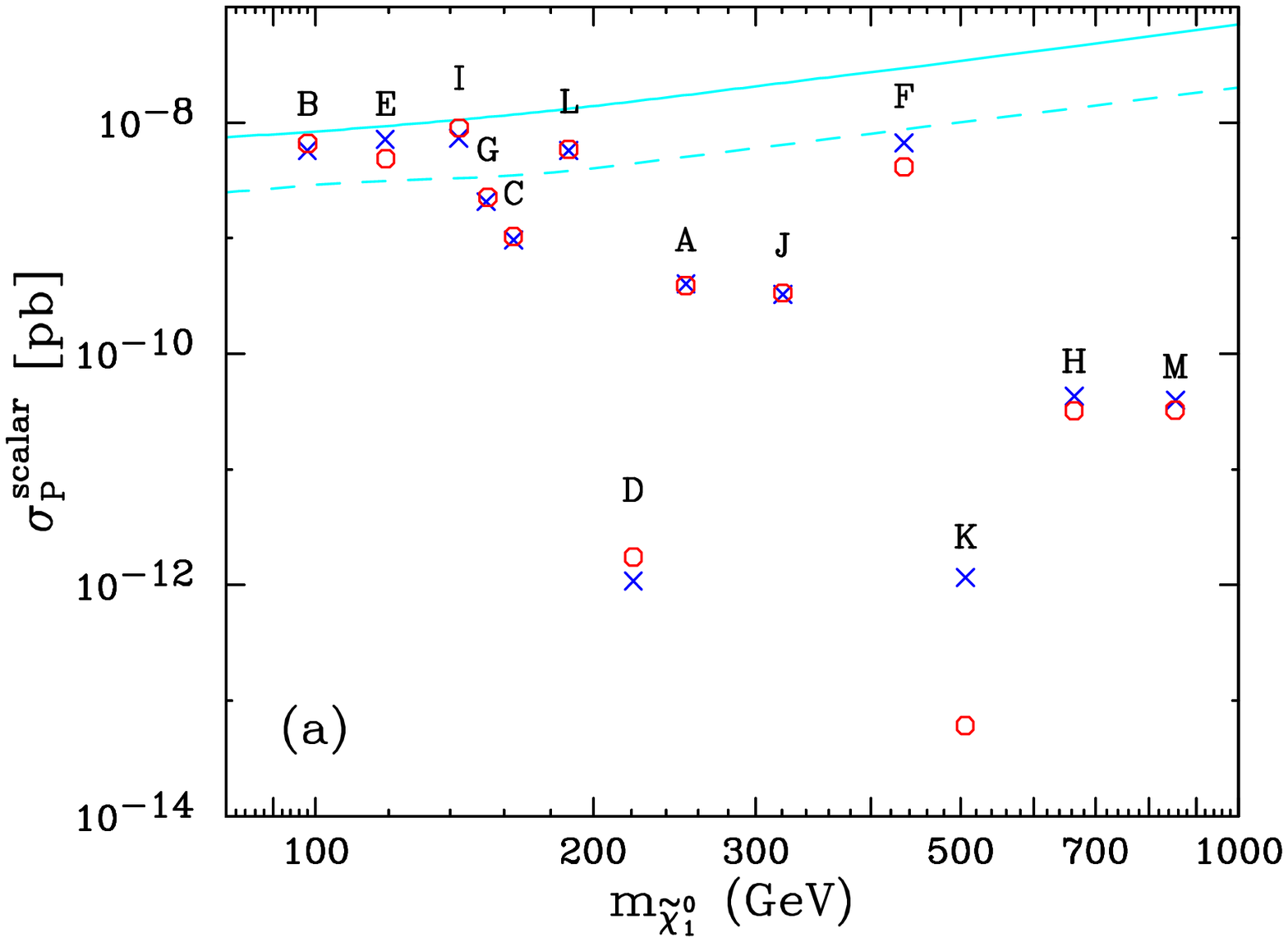}
\includegraphics[height=1.65in]{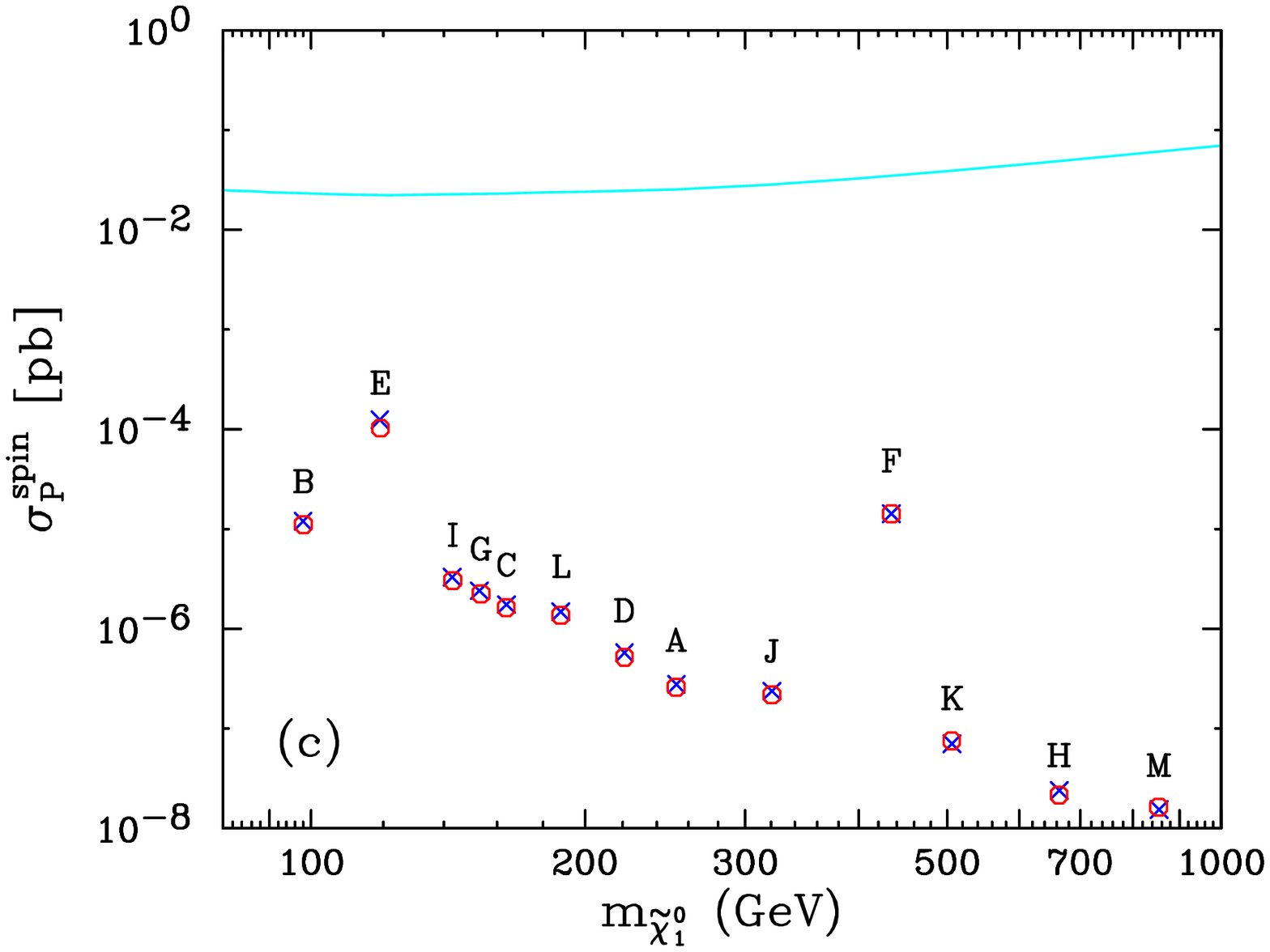}
\caption{\label{fig:DM}
{\it 
Elastic cross sections for (a)
spin-independent scattering and (b) spin-dependent scattering on
protons. Our predictions (blue crosses) 
are compared with those of  {\tt Neutdriver} \cite{Neutdriver} (red circles) for neutralino-nucleon
scattering.  Projected sensitivities (a) for CDMS
II~\cite{Schnee:1998gf} and CRESST~\cite{Bravin:1999fc} (solid) and
GENIUS~\cite{GENIUS} (dashed) and (b) for a 100 kg NAIAD
array~\cite{Spooner:2000kt} are also shown.
}}
\end{figure}

Because the LSP as dark matter is present locally, there are many
avenues for pursuing dark matter detection. Direct detection techniques
rely on an ample neutralino-nucleon scattering cross-section.
The prospects for direct detection for the benchmark points discussed
above \cite{EFFMO} are shown in Fig.~\ref{fig:DM}. 
This figure shows rates for the elastic
spin-independent  and spin dependent scattering cross sections of supersymmetric relics
on protons.
Indirect searches for supersymmetric dark matter via the products of
annihilations in the galactic halo or inside the Sun also have prospects
in some of the benchmark scenarios \cite{EFFMO}.

\begin{figure}[h]
\centering
\includegraphics[height=3.2in]{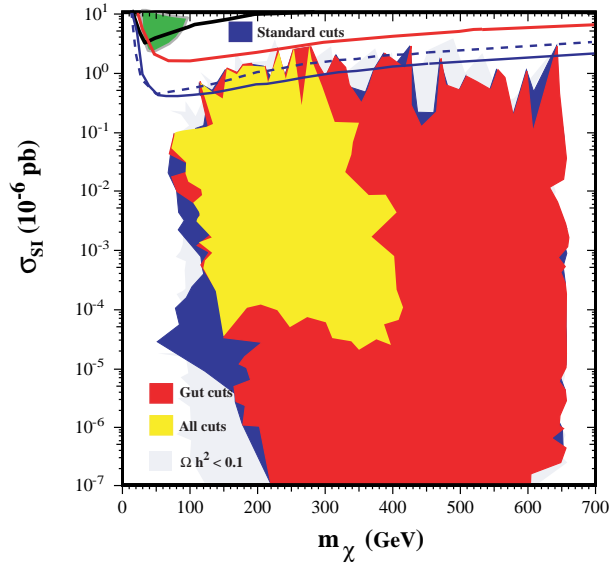}
\caption{\label{fig:Andyall}
{\it 
Ranges of the spin-independent 
cross section in the NUHM.
The ranges allowed by the  
cuts on $\ohsq$, $m_h$ and $b \to s \gamma$ have dark shading, those still 
allowed by the GUT stability cut have medium shading, and those still 
allowed after applying all the cuts including $g_\mu - 2$ have light 
shading.   The pale shaded region
corresponds to the extra area of points with low relic 
densities, whose cross sections have been rescaled appropriately. 
Also shown are the limits from the CDMS\protect\cite{cdms} and Edelweiss\protect\cite{edel}
experiments  as well as the recent CDMSII result \protect\cite{cdms2}
on the neutralino-proton elastic scattering cross section as a function
of the neutralino mass. The CDMSII limit is stronger than the 
Edelweiss limit which is stronger than the previous CDMS limit at higher $m_\chi$. 
The result reported by DAMA \protect\cite{dama} is found in the upper left.}}
\end{figure}

In Fig.~\ref{fig:Andyall}, we display the allowed ranges of 
the spin-independent  cross sections in the NUHM when we
sample randomly $\tan \beta$ as well as the other NUHM parameters \cite{efloso}. 
The raggedness of the boundaries of the shaded regions
reflects the finite sample size. The dark shaded regions includes
all sample points after the constraints discussed above (including the relic
density constraint) have been applied.  
In a random sample, one often hits points which are
are perfectly acceptable at low energy scales but
when the parameters are run to high energies approaching the GUT scale,
one or several of the sparticles mass squared runs negative \cite{fors}.  
This has been referred to as the GUT constraint here.
The medium shaded region
embodies those points after the GUT constraint has been applied.
After incorporating all
the cuts, including that motivated by $g_\mu - 2$, we find that the light shaded
region where the
scalar cross section has the range $10^{-6}$~pb $\ga
\sigma_{SI} \ga 10^{-10}$~pb, with
somewhat larger (smaller) values being possible in exceptional cases. If
the $g_\mu - 2$ cut is removed, the upper limits on the cross sections are
unchanged, but much lower values become possible: $\sigma_{SI} \ll
10^{-13}$~pb.  The effect of the GUT constraint on more general 
supersymmetric models was discussed in \cite{eoss3}.

The results from this analysis \cite{efloso} for the scattering cross section in the NUHM
(which by definition includes all CMSSM results)
are compared with the previous CDMS \cite{cdms} and Edelweiss \cite{edel}
bounds as well as the recent CDMSII results \cite{cdms2} in Fig.~\ref{fig:Andyall}. 
 While previous experimental
sensitivities were not strong enough to probe predictions of the NUHM, the current
CDMSII bound has begun to exclude realistic models and it is expected that 
these bounds improve by a factor of about 20.

  This work was partially supported by DOE grant
DE-FG02-94ER-40823.

%
%
%
%
%

%
%



\printindex
\end{document}